\documentclass[reprint, superscriptaddress, preprintnumbers, nofootinbib, amsmath, amssymb, aps, prd]{revtex4-2}
\usepackage{graphicx}
\usepackage{dcolumn}
\usepackage{bm}
\usepackage{xcolor}

\graphicspath{{Figure/}}
\usepackage[colorlinks, linkcolor=red,anchorcolor=green,citecolor=blue]{hyperref}

\usepackage{CJK}
\usepackage[normalem]{ulem}
\newcommand{\cjk}[1]{
\begin{CJK}{UTF8}{gbsn}{(#1)}\end{CJK}
}

\newcommand{\bth}{\boldsymbol{\vartheta}}
\newcommand{\mf}{\mathcal{F}}

\begin{document}
\preprint{RIKEN-iTHEMS-Report-25}
\title{
Towards constraining QCD phase transitions in neutron star interiors:\\
Bayesian Inference with TOV linear response analysis
}

\author{Ronghao Li\cjk{李荣浩}}
\email{lrh20@mails.tsinghua.edu.cn}
\affiliation{Department of Physics, Tsinghua University, Beijing 100084, China.}

\author{Sophia Han\cjk{韩君}}
\email{sjhan@sjtu.edu.cn}
\affiliation{Tsung-Dao Lee Institute, Shanghai Jiao Tong University, Shanghai 201210, China.}
\affiliation{School of Physics and Astronomy, Shanghai Jiao Tong University, Shanghai 200240, China.}

\author{Zidu Lin\cjk{林子都}}
\email{zlin23@utk.edu}
\affiliation{University of Tennessee, Knoxville, Tennessee 37996, USA.}

\author{Lingxiao Wang\cjk{王凌霄}}
\email{lingxiao.wang@riken.jp}
\affiliation{Interdisciplinary Theoretical and Mathematical Sciences Program (iTHEMS), RIKEN, Wako, Saitama 351-0198, Japan.}

\author{Kai Zhou\cjk{周凯}}
\email{zhoukai@cuhk.edu.cn}
\affiliation{School of Science and Engineering, The Chinese University of Hong Kong, Shenzhen (CUHK-Shenzhen), Guangdong, 518172, China.}
\affiliation{Frankfurt Institute for Advanced Studies, Ruth Moufang Strasse 1, D-60438, Frankfurt am Main, Germany.}

\author{Shuzhe Shi\cjk{施舒哲}}
\email{shuzhe-shi@tsinghua.edu.cn}
\affiliation{Department of Physics, Tsinghua University, Beijing 100084, China.}
\affiliation{State Key Laboratory of Low-Dimensional Quantum Physics, Tsinghua University, Beijing 100084, China.}

\date{\today}
\begin{abstract}
The potential hadron-to-quark phase transition in neutron stars has not been fully understood as the property of cold, dense, and strongly interacting matter cannot be theoretically described by the first-principle perturbative calculations, nor have they been systematically measured through terrestrial low-to-intermediate energy heavy-ion experiments.
Given the Tolman--Oppenheimer--Volkoff (TOV) equations, the equation of state (EoS) of the neutron star (NS) matter can be constrained by the observations of NS mass, radius, and tidal deformability. 
However, large observational uncertainties and the limited number of observations currently make it challenging to strictly reconstruct the EoS, especially to identify interesting features such as a strong first-order phase transition.
In this work, we study the dependency of reconstruction quality of the phase transition on the number of NS observations of mass and radius as well as their uncertainty, based on a fiducial EoS.
We conquer this challenging problem by constructing a neural network, which allows one to parametrize the EoS with minimum model-dependency, and by devising an algorithm of parameter optimization based on the analytical linear response analysis of the TOV equations. This work may pave the way for the understanding of the phase transition features in NSs using future $x$-ray and gravitational wave measurements.
\end{abstract}

\maketitle
\section{Introduction}
\label{sec:intro}

The study of the equation of state (EoS) for cold dense quantum chromodynamics (QCD) matter has long posed a significant challenge in nuclear physics. On one hand, theoretical computations of the cold dense matter EoS from first-principle lattice QCD calculations are prohibited due to the sign problem (see~\cite{Aarts:2015tyj, deForcrand:2009zkb} for recent reviews). 
On the other hand, nuclear matter generated in terrestrial heavy-ion collision experiments typically resides in the high-temperature regime ($T\gtrsim 100~\mathrm{MeV}/k_\mathrm{B}$)~\cite{Dexheimer:2020zzs, Fukushima:2020yzx, Sorensen:2023zkk}.
In contrast, mature neutron stars (NSs) exhibit low temperatures $T \sim 100\,\mathrm{eV}/k_\mathrm{B}$ but high baryon number densities $n \sim 0.3~\mathrm{fm}^{-3}$ in their interiors, offering a promising and unique avenue for probing the thermodynamic properties of cold dense QCD matter~\cite{Lattimer:2012nd, Baym:2017whm}. 
With rapid advancements in gravitational-wave astrophysics and multi-messenger astronomy, particularly since pioneering observations from the advanced gravitational-wave detector network LIGO/Virgo/KAGRA~\cite{LIGOScientific:2018cki, LIGOScientific:2017vwq, LIGOScientific:2020aai, LIGOScientific:2020zkf} and the \textit{Neutron Star Interior Composition ExploreR} (NICER) mission~\cite{Miller:2019cac, Riley:2019yda, Riley:2021pdl, Miller:2021qha, Reardon:2015kba}, both the quantity and precision of NS observational data with respect to constraining dense matter EoS have significantly improved in recent years~\cite{Yunes:2022ldq}. 

Assuming the EoS for cold dense matter is known, one can solve the renowned Tolman--Oppenheimer--Volkoff (TOV) equations~\cite{Tolman:1939jz, Oppenheimer:1939ne} in a static and spherically symmetric frame to derive the relationship between the NS masses ($M$) and radii ($R$), with the central pressure acting as the latent variable. In essence, the TOV equations function as a mapping from the EoS to the $M$-$R$ relation. 
As shown by Lindblom~\cite{1992ApJ...398..569L}, this functional mapping is reversible, meaning that the EoS can be precisely reconstructed if comprehensive information of the full $M$-$R$ curve is given. 
In reality, however, only limited observations with sizable uncertainties of the NS mass, radius, and tidal deformability are available at the current stage, preventing concrete determination of the EoS by possibly inverting the $M$-$R$ relation through TOV equations.
In this context, there have been so far two plausible approaches: first, the functional mapping between the EoS and the $M$-$R$ relation can be parametrized by some empirical formulae that fit at least the most commonly used EoSs based on physical models~\cite{Ofengeim:2023nvc, Sun:2023xkg, Sun:2024nye}; second, one may employ Bayesian inference to infer the underlying EoS, utilizing constraints from gravitational-wave and electromagnetic observations~\cite{Huth:2020ozf, Huth:2021bsp, Pang:2021jta, Chatziioannou:2024tjq, Al-Mamun:2020vzu}. 

The EoS, $\varepsilon(P)$, relating the pressure and the energy density of dense matter is a fundamental thermodynamic property of strongly-interacting QCD, which possibly exhibits first-order phase transitions (PTs) within density regimes relevant for NS interiors, for example the crust-core liquid-gas PT~\cite{Baym:1975mf, 1983Natur.305..410S, Douchin:2001sv}, PTs from nucleonic matter to hyperonic matter~\cite{Schaffner-Bielich:2008zws, Vidana:2013nxa, Schaffner-Bielich:2000nft}, and possible first-order PT from nuclear or hadronic matter to quark matter~\cite{Stephanov:2006zvm, Annala:2019puf, Bzdak:2019pkr,Dexheimer:2020zzs, Annala:2023cwx}. 
The last one, in particular, is expected to result in a critical endpoint on the QCD phase diagram, marking the switching from a smooth crossover to a first-order PT~\cite{Dexheimer:2020zzs, Stephanov:2006zvm, Bzdak:2019pkr}.
Nevertheless, with current limitations on the amount of observational data and the sizable uncertainties associated with them, it is widely acknowledged that confidently reconstructing the detailed features of the NS EoS --- particularly the hadron-to-quark PT --- remains elusive~\cite{Essick:2023fso, Lin:2023cbo, Annala:2023cwx}. This naturally raises the question:
\textit{How many $M$-$R$ observations are needed, and how precise must future observations be to confirm the existence and characterize the features of a potential PT in NS inner cores?}

To address this question quantitatively, one must infer the EoS based on a sufficiently flexible assumption (i.e., to perform a \textit{nonparametric} Bayesian analysis) while ensuring that the update process remains efficient for handling numerous observations. Specifically, the EoS should be represented by a flexible enough functional form capable of capturing the full complexity of the system, e.g., those parametrized with sufficient number of parameters. 
This approach guarantees that, in the ideal scenario with sufficient amount of high-precision NS observations, the EoS can be reconstructed with high fidelity, capturing intricate details such as a first-order PT. 
While Gaussian Processes (GPs) are another prominent example of nonparametric methods that provide flexible, probabilistic models, we opt for a deep neural network (DNN) representation of the EoS, given their scalability and ability to handle complex, high-dimensional parameter spaces efficiently, which is crucial for the large-scale nature of our analysis.
A nonparametric Bayesian analysis is then only feasible if one can update the involved large set of parameters with high efficiency. 

Gradient-based optimization has proven successful in parameter-intensive tasks, so the gradients of NS observables with respect to EoS parameters can be leveraged for this purpose.
Such a numerical tool has been partially developed in previous works by some of the authors~\cite{Soma:2022qnv, Soma:2022vbb}, where the EoS is represented by a DNN, and the mapping between the EoS and the $M$-$R$ relation (i.e., the TOV equations) is approximated by another DNN. This setup enables numerical computation of EoS parameter gradients via auto-differentiation. However, the TOV-equation-approximator DNN is not guaranteed to accurately reproduce the EoS-to-($M$-$R$) mapping, and quantifying the associated systematic uncertainties remains nontrivial. 
In this work, we resolve this issue by avoiding the use of the TOV-equation-approximator DNN. Instead, we solve the TOV equations directly and compute the derivatives with respect to changes in the EoS in a numerically efficient manner. 
This method, also referred to as \textit{physics-driven learning}, has been applied to the study of other inverse problems in nuclear physics~\cite{Karpie:2019eiq, NNPDF:2021uiq, Gao:2022iex, Shi:2021qri, Wang:2021jou, Shi:2022yqw, Li:2022ozl, Li:2025csc, Luo:2024iwf, Wang:2024ykk, Wang:2024dzc, Wang:2024bpl}. For recent reviews, see~\cite{Zhou:2023pti, Aarts:2025gyp}.

In this work, we develop the computational method and perform a proof-of-principle study, 
based on which the aforementioned question could eventually be answered.
With newly developed tool introduced in Sec.~\ref{sec:algorithm}, we systematically study the degradation of the reconstruction quality of a fiducial EoS in Sec.~\ref{sec:results}, by gradually increasing the uncertainty of data points on its corresponding $M$-$R$ curve. The ``lower limit" of the $M$-$R$ observations needed to verify the PT of this fiducial EoS will also be discussed.
Following the summary and discussion in Sec.~\ref{sec:summary}, we provide appendices for a numerical validation of the derivative analysis (\ref{app:validation}) and for details of importance sampling (\ref{app:importance_sampling}), respectively.

\section{Description of the Algorithm}
\label{sec:algorithm}

The key to reconstructing the EoS from NS observables in a nonparametric manner -- especially considering discrete and noisy observational data -- is to perform Bayesian Inference with an arbitrary large number of parameters. This can be achieved if one can efficiently compute the derivative of the posterior distribution with respect to the model parameters. Noting that the posterior distribution takes NS observables as input, which depends on the EoS implicitly, computing the parameter gradients is not straightforward. This section provides a detailed description of our algorithm that computes the parameter gradients based on a linear response analysis of the TOV equations. The gradient is useful not only in guiding the direction of parameter update iterations, but also in physics analysis of the sensitivity of NS observables with respect to changes in the EoS.

In this work, we use the following notations:
\begin{itemize}
\item The index $i = 1, \cdots, N_\text{obs}$, labels the $i$-th NS.
\item The index $j = 1, \cdots, N_\text{points}$, labels the $j$-th point of the discrete EoS representation.
\item The index $k = 1, \cdots, N_\text{prm}$, labels the $k$-th parameter in representing the EoS.
\item We use the curly bracket $\{_n\}$ as a shorthand of the list $\{X_1, \cdots, X_n, \cdots, X_{\max} \}$, where $n$ can be one of the indices listed above, and $X$ can be any relevant physical quantity. If more than one indices appear in the bracket, we denote $\{X_{n,m}\}_n$ to indicate that the list runs over $n$-indices.
\end{itemize}

\subsection{Likelihood distribution of the Equation of State}
\label{sec:algorithm:general}

In order to reconstruct the EoS, with uncertainties reflecting the finite-precision in astronomical measurements, we obtain the posterior distribution of the reconstructed EoS by invoking Bayesian Analysis (BA) --- a statistical method that is frequently used in data-driven studies. See Refs.~\cite{steiner:2013neutron, ozel:2016masses, Bernhard:2016tnd, Bernhard:2019bmu, JETSCAPE:2020shq, JETSCAPE:2020mzn, JETSCAPE:2021ehl, Heffernan:2023gye, Heffernan:2023utr} for examples of revealing QCD thermodynamical properties in NS physics and relativistic heavy-ion collisions. The Likelihood function, which is central to BA, can be computed by the $\chi^2$-function, 
\begin{equation}
    L(\bth|\text{data}) \propto \exp(-\chi^2(\bth)/2),
\end{equation} 
where the vector $\bth$ is a shorthand of the unknown variables in the model. They are the central pressures of the NSs of interest and all the parameters for representing the EoS, including the critical pressure and latent heat of the possible first-order phase transition(s). In practice, the $\chi^2$-function is the sum of independent measurements of different NSs, 
\begin{align}
    \chi^2(\bth) = \sum_i \chi^2_i(\bth),
\end{align}
where each $\chi^2_i(\bth)/2$ is logarithm of the posterior distribution of the NS observables by taking the reconstructed values $(M_i(\bth), R_i(\bth))$. While $\chi^2_i(\bth)$ is usually complex, it takes a simple form for an ideal NS with uncorrelated Gaussian measurement of the mass ($\overline{\text{M}}_i \pm \Delta_{M,i}$) and the radius ($\overline{\text{R}}_i \pm \Delta_{R,i}$), in which one can naturally define the $\chi^2$-function as the uncertainty weighted distances between the reconstructed values and the corresponding measurements, 
\begin{align}
\chi^2_i(\bth)=\;&
    \sum_{O\in\{M, R\}}\frac{(O_{i} - \overline{\text{O}}_{i})^2}{\Delta_{O,i}^2}\,. 
\label{eq:chi_sq_Gaussian}
\end{align}
As noted before, $i$ is the index of the NS. In addition to the NS observables, other information of the EoS can be encoded as the Prior distribution, $\mathrm{Prior}(\bth)$, and the final Posterior distribution of all parameters is given by 
\begin{align}
    \mathrm{P}(\bth) = \frac{\mathrm{Prior}(\bth)\, 
    e^{-\frac{\chi^2(\bth)}{2}}}{\mathcal{N}_\mathrm{E}}\,,
\end{align}
with $\mathcal{N}_\mathrm{E}$ being the evidence which ensures the integration of $\mathrm{P}(\bth)$ to be unity.

By finding the parameter set that maximizes the posterior,
\begin{align}
    {\bth}^\mathrm{opt} \equiv\;& \mathrm{arg\,max}_{\bth} \,\mathrm{P}(\bth),
\end{align}
one obtains the optimal EoS that best matches the astronomical observations taking into account prior knowledge, which is also referred to as the Maximum a Posteriori (MAP). For later convenience, we denote $\mathcal{R}(\bth) \equiv -\ln\mathrm{Prior}(\bth)$ as the regulator defined by Prior distribution and $\mathcal{J}(\bth)\equiv \frac{1}{2}\chi^2(\bth) + \mathcal{R}(\bth)$ as the total loss function. 

Performing BA in a high-dimensional parameter space (with $N_\mathrm{prm} \sim 10^{2-3}$), even if only searching for the MAP, is generally challenging. 
In statistics, Markov-Chain Monte-Carlo (MCMC)~\cite{1_MCMC} is usually invoked to approach the region around ${\bth}^\mathrm{opt}$ and construct the Posterior distribution. 
At each step of the MCMC sampling, a new parameter set is proposed randomly, and its acceptance probability is designed such that the desired likelihood function can be approached. As the acceptance rate decreases rapidly with the increasing number of parameters, the MCMC procedure is only applicable to problems with a few parameters; for those with $N_\mathrm{prm} \gtrsim 100$, MCMC usually fails to converge within reasonable computation time.

Nevertheless, if the loss function is differentiable with respect to the model parameters, one may obtain ${\bth}^\mathrm{opt}$ according to gradient-based methods, which is applicable to problems with huge amount of parameters. Fortunately, our problem of interest falls into such a category. 
This is not obvious because $ \chi^2 $ is the function of the NSs' masses, radii, and tidal deformabilities, all of which depend on the EoS implicitly. While one cannot compute the derivative of implicit functions in general, we perform a linear response analysis to the TOV equations and obtain the formulae of changes in mass and radius with respect to an arbitrary perturbation in the EoS, or in the central pressure of a NS. Thus, the parameter gradient of $\chi^2$ can be computed. 
With $\nabla_{\bth}\chi^2$, one may update $\bth$ iteratively to minimize $\chi^2$, i.e., maximize the likelihood function, and explore the region around that. By doing so, one may propose parameter sets most efficiently to perform a BA.

\subsection{Variation analysis of NS observables}
\label{sec:algorithm:linear}

\subsubsection{Preparation: change of variables in the TOV equations}

The TOV equations~\cite{Tolman:1939jz, Oppenheimer:1939ne} describe stable configurations of NSs that are spherically symmetric, giving the pressure [$P(r)$] at a distance $r$ from the stellar center as well as the mass enclosed by the corresponding spherical shell, labeled by $m(r)$. 
The former is governed by the hydrostatic equilibrium in general relativity that the attractive gravitational force is balanced by the gradient of internal pressure,
\begin{equation}
    \frac{\mathrm{d}P}{\mathrm{d}r}  =  -\frac{(P + \varepsilon ) (m+ 4 \pi r^3 P)}{r^2 - 2m r}\,,
    \label{eq:TOV_P}
\end{equation}
and the latter is ruled by the continuity equation
\begin{equation}
    \frac{\mathrm{d}m}{\mathrm{d}r} = 4 \pi r^2 \varepsilon\,, 
     \label{eq:TOV_m}
\end{equation}
where $\varepsilon$ is the energy density determined by the pressure through the underlying microscopic equation of state,
\begin{equation}
    \varepsilon   =   \varepsilon(P)\,.
     \label{eq:TOV_e}
\end{equation}
Throughout this paper, we use natural units with $c = G = \hbar = 1$. It shall be worth noting that the rotation effects are negligible for NSs with frequency $\lesssim 0.1 \mathrm{kHz}$~\cite{Haensel:2009wa, Haensel:2016pjp, Suleimanov:2020ijb}, and therefore the TOV equations, which take a static and spherical symmetric metric, are feasibly applicable to slow rotating NSs. 
Meanwhile, it has been found that the gravitational corrections on the EoS can be safely neglected, see e.g.~\cite{Li:2022url}.

Given the central pressure $P(r=0)=P_c$ as well as the initial condition for mass $m(r=0)=0$, one can solve the differential equations for $P(r)$ and $m(r)$ with increasing $r$ starting from the center and progressing outward, until the pressure decreases to zero, $P(r)=0$. The final-state radius, labeled as $R$, defines the radius of a NS, and the enclosed mass is the total mass of the NS, $M=m(R)$.
In practice, the boundary (surface) pressure is usually taken as a nonzero but small enough value, and we denote such a value as $P_\text{bnd}$.

In addition to the NS mass and radius, tidal deformability is another macroscopic observable that is sensitive to the nuclear EoS~\cite{Flanagan:2007ix, Damour:2009vw, Hinderer:2007mb, Postnikov:2010yn}. It characterizes the response relationship of the induced mass-quadrupole moment of a NS to an external perturbing tidal field (the gravitational field of a companion star), which can be inferred from gravitational wave detections of binary NS merger events. 
Although our examples of the EoS reconstruction presented later in this work only involve information of mass and radius, we also perform linear response analysis for the tidal deformability for reference.

Denoting the compactness parameter $\beta \equiv M/R$, the dimensionless tidal deformability is given by~\cite{Flanagan:2007ix, Damour:2009vw, Hinderer:2007mb, Postnikov:2010yn}:
\begin{align}
\begin{split}
\Lambda \equiv\;&
    \frac{\lambda}{M^5} 
    =  \frac{2}{3}\frac{k_2}{\beta^5}
\\=\;&  
    \frac{16}{15}(1-2\beta)^2 \big(2-2\beta+(2\beta-1)Y\big)
\\&\times
    \Big((3-12\beta+13\beta^2-2\beta^3+2\beta^4)4\beta
\\&
    +(-3+15\beta-22\beta^2+6\beta^3+4\beta^4)2\beta Y
\\&
    +3(1-2\beta)^2\big(2-2\beta+(2\beta-1)Y\big)\ln(1-2\beta)\Big)^{-1}\,,
\end{split}
\end{align}
where $Y\equiv y(R)$ is the boundary value of function $y(r)$. The latter follows the differential equation:
\begin{align}
\begin{split}
r\frac{\mathrm{d}y}{\mathrm{d}r} 
    =\;&
    - y^2
    -\frac{1 + 4\pi r^2 (P-\varepsilon)}{1 - 2\,m/r} y
    + \frac{4(m/r + 4\pi r^2 P)^2}{(1 - 2\,m/r)^2}
\\&
    +
    \frac{6 - 4\pi r^2 (5\varepsilon+9P+(\varepsilon+P) \kappa_s)}{1 - 2\,m/r}\,,
\end{split}
\label{eq:TOV_y}
\end{align}
with the boundary condition $y(r=0)=2$, and
\begin{align}
    \kappa_s \equiv c_s^{-2} = \frac{\mathrm{d}\varepsilon}{\mathrm{d}P}
\end{align}
is the inverse speed of sound squared.\footnote{Suppose one takes $\Lambda$ into account in the reconstruction, the sum of observables in Eqs.~\eqref{eq:chi_sq_Gaussian} and \eqref{eq:correlation_kernal} shall run over $\{M,R,\Lambda\}$.}

In general, one solves the TOV and tidal equations~(\ref{eq:TOV_P}), (\ref{eq:TOV_m}), (\ref{eq:TOV_y}) using the radius $r$ as the independent variable. However, this makes it numerically inefficient and analytically hard to analyze the linear response of the NS observables against the perturbation in the EoS and/or in the central pressure. Instead, it would be more convenient to solve the equations with respect to the pressure.
Noting that the pressure varies by several orders of magnitude from the NS center to its surface, we take the log-pressure
\begin{align}
    \xi\equiv \ln\frac{P}{P_\mathrm{bnd}}
    \label{eq:xi}
\end{align}
as the independent variable. Here, $P_\mathrm{bnd}$ is a low-enough pressure that defines the boundary (surface) of a NS. In practice, we take $P_\mathrm{bnd} = 1.4 \times 10^{-12} \,\mathrm{MeV}/\mathrm{fm}^3$, and we have checked that $M$, $R$, and $\Lambda$ do not change when we vary $P_\mathrm{bnd}$ by a few orders of magnitude. 
Meanwhile, instead of $r$, we define $v\equiv r^3$ as the variable to be solved to ensure numerical stability. 
We solve the differential equation set from $\xi = \xi_c\equiv\ln\frac{P_{c}}{P_\mathrm{bnd}}$ to $\xi=0$, i.e., from $P=P_c$ to $P=P_\text{bnd}$ according to uniform step in $\xi$. 
With change of variables, the TOV and tidal equations~(\ref{eq:TOV_P}), (\ref{eq:TOV_m}), (\ref{eq:TOV_y}) now become
\begin{align}
\begin{split}
\frac{\mathrm{d}v}{\mathrm{d}\xi} =\;&
   - \mathcal{K}_v  \,, 
   \qquad\qquad
\frac{\mathrm{d}m}{\mathrm{d}\xi} =
   - \mathcal{K}_m\,,\\
\frac{\mathrm{d}y}{\mathrm{d}\xi} =\;&     
    -\frac{\mathcal{K}_v}{3v}r\frac{\mathrm{d}y}{\mathrm{d}r} 
    = - \mathcal{K}_{y}\,,
\end{split}
   \label{eq.tov_P_m}
\end{align}
where
\begin{align}
\mathcal{K}_v \equiv \;&
    \frac{3(v^{\frac{1}{3}} - 2\,m)}{(m/v+4\pi P)(1+\varepsilon/P)} \,,\\
\mathcal{K}_m \equiv \;&
    \frac{4\pi \varepsilon}{3}\, \mathcal{K}_v\,,\\
\begin{split}
\mathcal{K}_{y} \equiv\;& 
    -\frac{\mathcal{K}_v}{3v} y^2 
    -\frac{v^{-\frac{2}{3}} + 4\pi (P-\varepsilon)}{(m/v+4\pi P)(1+\varepsilon/P)}\,y
\\&
    -\frac{4\pi (5\varepsilon+9P+(\varepsilon+P) \kappa_s)-6 v^{-\frac{2}{3}}}{(m/v+4\pi P)(1+\varepsilon/P)}
\\&
    + \frac{4m + 16\pi P v}{(v^{\frac{1}{3}} - 2\,m)(1+\varepsilon/P)}\,.
\end{split}\label{eq:Ky}
\end{align}

In the presence of first-order phase transition, the inverse speed of sound could contain Dirac $\delta$-function pulses. 
For instance, suppose there exists a phase transition point at $\xi=\xi_\mathrm{PT}$, then the inverse speed of sound squared contains the regular part and a pulse 
\begin{align}
\kappa_s(\xi) =\;& 
    \frac{\mathrm{d}\varepsilon}{\mathrm{d}P}\Big|_{\xi\neq\xi_\mathrm{PT}} + \frac{\Delta \varepsilon}{P(\xi_\mathrm{PT})} \delta(\xi-\xi_\mathrm{PT})\,.
\end{align}
It leads to a $\delta$-function in the $y$-kernel~\eqref{eq:Ky}, which results in a jump in $y$ for the log-pressure just below ($\xi_\mathrm{PT}^-$) and above ($\xi_\mathrm{PT}^+$) the phase transition point~\cite{Postnikov:2010yn, Takatsy:2020bnx},
\begin{align}
    y(\xi_\mathrm{PT}^-) - y(\xi_\mathrm{PT}^+) = 
    -\frac{\Delta\varepsilon}{\frac{m}{4\pi v}+P}\,.
\end{align}

We solve Eq.~\eqref{eq.tov_P_m} with initial conditions given by the asymptotic behaviors near center ($\xi \to \xi_c$, or $P \to P_c$),
\begin{align}
    v(\xi) =\;& \bigg(\frac{P_c-P}{2\pi(P_c+\varepsilon_c)(P_c+\varepsilon_c/3)}\bigg)^{3/2}\,,\\
    m(\xi) =\;& \frac{4\pi}{3} \varepsilon(\xi_c)\, v(\xi)\,,\\
    y(\xi)=\;&2 + \mathcal{O}(\xi-\xi_c)\,,
\end{align}
and obtain physical observables at the outer edge ($\xi \to 0$, or $P \to P_\text{bnd}$),
\begin{align}
    M = m(\xi=0)\,,\;\;
    R = v^{\frac{1}{3}}(\xi=0)\,,\;\;
    Y = y(\xi=0)\,.
\end{align}
For later convenience, we denote that
\begin{align}
\mathcal{K}_{O,\varphi} \equiv 
    \frac{\partial \mathcal{K}_O}{\partial \varphi}\,,
    \label{eq.K_dev}
\end{align}
for $O\in\{v,m,y\}$ and $\varphi\in\{P,\varepsilon,\kappa_s,v,m,y\}$. Particularly, $\mathcal{K}_{m,y} = \mathcal{K}_{v,y} = \mathcal{K}_{m,\kappa_s} = \mathcal{K}_{v,\kappa_s} = 0$. It should be noted that $\varepsilon$, $\kappa_s$, and $P$ are regarded as independent variables when performing the partial derivatives.

With the preparation of the change of variables, we are ready to derive the linear response of NS observables against changes in the EoS and in the central pressure --- 
they will be discussed in Sec.~\ref{sec:algorithm:linear:eos} and Sec.~\ref{sec:algorithm:linear:preussure}, respectively.

\subsubsection{Perturbation in the EoS}
\label{sec:algorithm:linear:eos}

Given an arbitrary perturbation in the EoS, $\varepsilon(\xi) \to \varepsilon(\xi) + \Delta_\varepsilon(\xi)\,\delta_\varepsilon$, where $\delta_\varepsilon$ is a small parameter ensuring the applicability of the linear perturbation theory, we denote the changes in the observables as $v(\xi) \to v(\xi) + \Delta_v(\xi)\,\delta_\varepsilon$, $m(\xi) \to m(\xi) + \Delta_m(\xi)\,\delta_\varepsilon$, and $y(\xi) \to y(\xi) + \Delta_y(\xi)\,\delta_\varepsilon$. Note that the change in the inverse sound speed squared is given by $\kappa_s(\xi) \to \kappa_s(\xi) + \frac{\mathrm{d}\Delta_\varepsilon(\xi)}{P\,\mathrm{d}\xi}\delta_\varepsilon$, correspondingly. Keeping up to linear perturbations, the changes of the equations of motion~\eqref{eq.tov_P_m} read
\begin{align}
\begin{split}
    -\frac{\mathrm{d}\Delta_v}{\mathrm{d}\xi} 
    =\;& \mathcal{K}_{v,v}\Delta_v 
    + \mathcal{K}_{v,m}\Delta_m 
    + \mathcal{K}_{v,\varepsilon}\Delta_\varepsilon\,,
\end{split}
\label{eq:eom:dv_de_diff}\\
\begin{split}
    -\frac{\mathrm{d}\Delta_m}{\mathrm{d}\xi} 
    =\;& \mathcal{K}_{m,v}\Delta_v 
    + \mathcal{K}_{m,m}\Delta_m 
    + \mathcal{K}_{m,\varepsilon}\Delta_\varepsilon\,,
\end{split}
\label{eq:eom:dm_de_diff}\\
\begin{split}
    -\frac{\mathrm{d}\Delta_y}{\mathrm{d}\xi} 
    =\;& \mathcal{K}_{y,v}\Delta_v 
    + \mathcal{K}_{y,m}\Delta_m 
    + \mathcal{K}_{y,y}\Delta_y 
\\&
    + \Big(\mathcal{K}_{y,\varepsilon}
    + \frac{\mathcal{K}_{y,\kappa_s}}{P}
    \frac{\mathrm{d}}{\mathrm{d}\xi} \Big)\Delta_\varepsilon\,.
\end{split}
\label{eq:eom:dy_de_diff}
\end{align}

Given a perturbation $\Delta_\varepsilon(\xi)$, one may solve Eqs.~(\ref{eq:eom:dv_de_diff}--\ref{eq:eom:dy_de_diff}) and obtain the changes of observables $\Delta_v(\xi)$, $\Delta_m(\xi)$, $\Delta_y(\xi)$. 
Suppose we perform a Dirac-$\delta$ function perturbation at $\xi=\xi'$ in the energy density, $\Delta_\varepsilon(\xi) = \delta(\xi-\xi')$, and the corresponding changes in observables are denoted as $\Delta_v(\xi|\xi')$, $\Delta_m(\xi|\xi')$, and $\Delta_y(\xi|\xi')$, respectively. It is not hard to show that responses to different perturbations in $\varepsilon$ are linearly addable. That is, if one introduces a perturbation in the energy density as $\Delta_\varepsilon(\xi) = \sum_k e_k\, \delta(\xi-\xi'_k)$, the change of observables can be given by $\Delta_O(\xi) = \sum_k e_k\, \Delta_O(\xi|\xi'_k)$. The same argument holds when generalizing the summation of discrete modes [$\sum_k e_k f(\xi_k')$] to the integration of continuous ones [$\int \mathrm{d}\xi'\, e(\xi') f(\xi')$]. Note that any physical variations on an EoS can be expressed as an integration of Dirac-$\delta$ functions,
\begin{align}
\begin{split}
    \Delta_\varepsilon(\xi) = \int \Delta_\varepsilon(\xi')\, \delta(\xi-\xi') \,\mathrm{d}\xi'\,,
\end{split}
\label{eq.eos_sum_perturbation}
\end{align}
and it would lead to the perturbation in $O$ as
\begin{align}
\Delta_O(\xi) =\;&
    \int_\xi^{\xi_c} \Delta_\varepsilon(\xi')\, \Delta_O(\xi|\xi')\, \mathrm{d}\xi'\,.
\end{align}

\begin{figure}[!hbtp]\centering
    \includegraphics[width=0.45\textwidth]{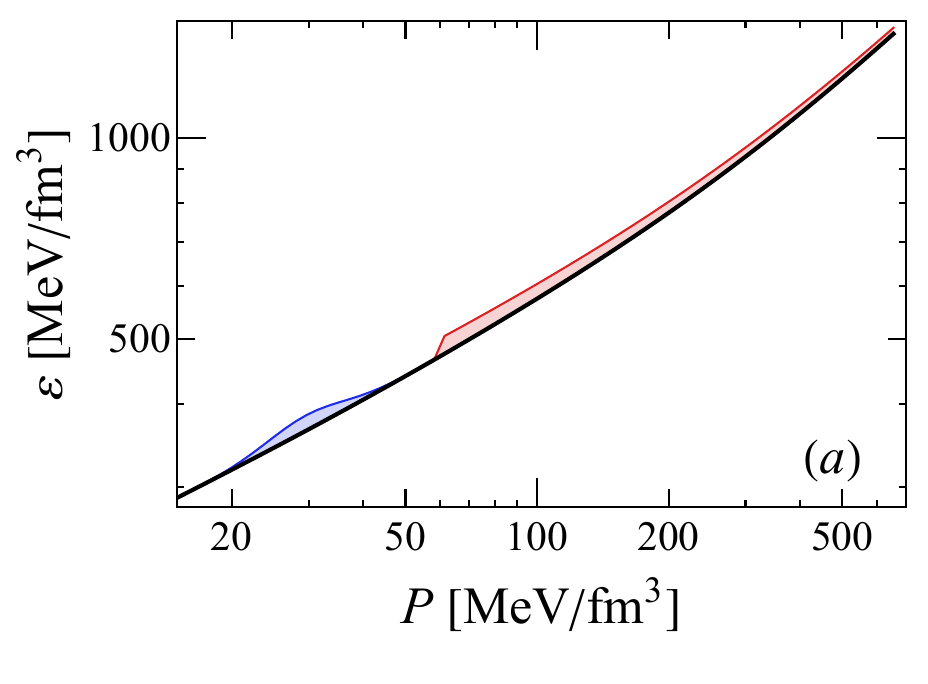}
    \includegraphics[width=0.45\textwidth]{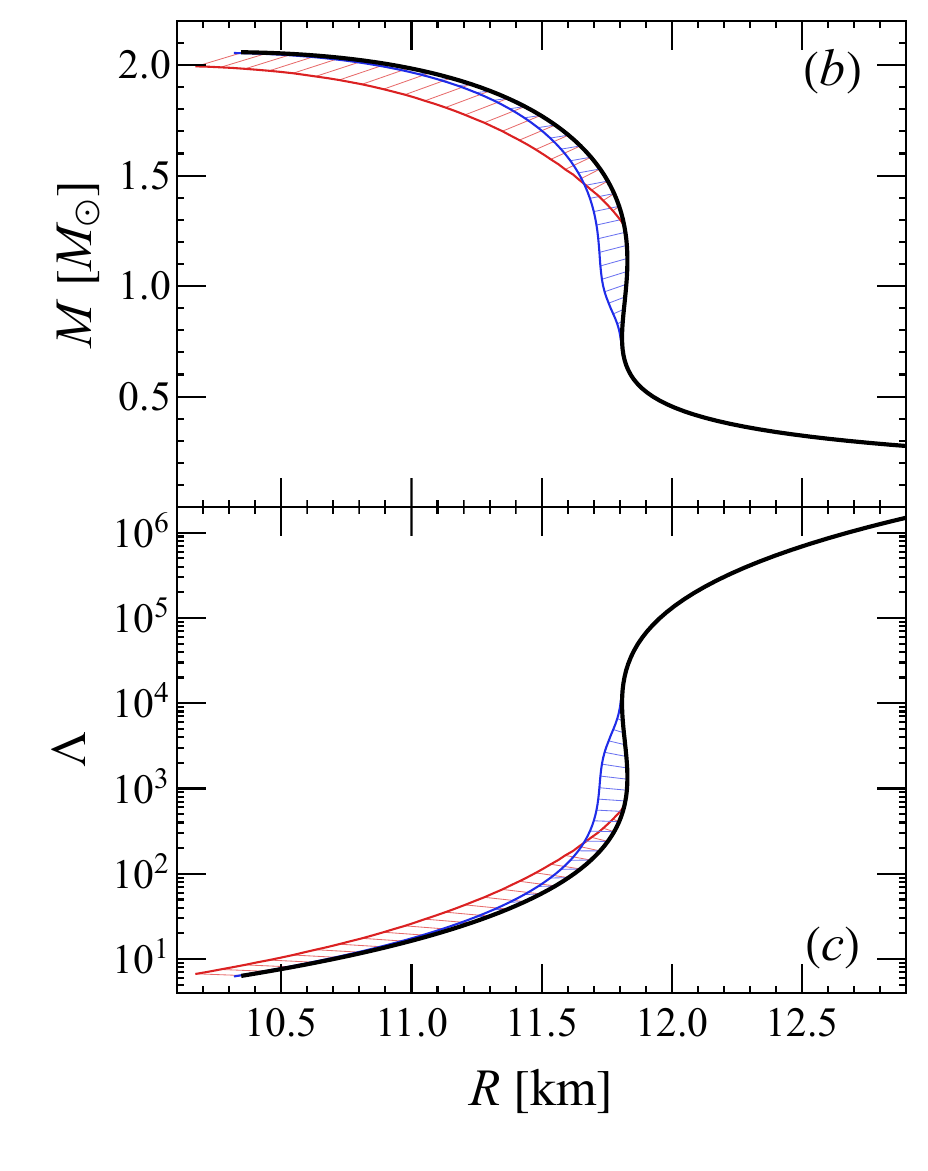}
    \caption{
    NS equations of state ($a$) and their corresponding $M$-$R$ ($b$) and $\Lambda$-$R$ ($c$) curves. Black curves correspond to the \texttt{SFHo} EoS, while red and blue curves include a step function and a Gauss function perturbation on top of that, respectively. The short colored lines connect the $M$, $R$, and $\Lambda$ values of NSs with the same central pressure before and after the perturbation.}
\label{fig:illustration}
\end{figure}

Finally, given an arbitrary parametrization of the EoS, $\varepsilon=\varepsilon(\xi|\boldsymbol{\vartheta})$, perturbation in parameters would lead to a change $\varepsilon(\xi) \to \varepsilon(\xi) + \sum_k \frac{\partial\varepsilon(\xi|\boldsymbol{\vartheta})}{\partial \vartheta_k} \delta \vartheta_k$. Therefore,
\begin{align}
    \frac{\delta O(\xi)}{\delta \vartheta_k} = \int_\xi^{\xi_c} \frac{\partial\varepsilon(\xi|\boldsymbol{\vartheta})}{\partial \vartheta_k}\, \Delta_O(\xi|\xi')\, \mathrm{d}\xi'\,.
    \label{eq:eom:arb_gradient}
\end{align}

For illustrative purposes, we show in Fig.~\ref{fig:illustration} how the NS mass, radius, and tidal deformability respond to the change in the EoS. On top of the baseline \texttt{SFHo} EoS, we respectively introduce i) a step function perturbation, and ii) a Gauss function perturbation, and then compute the induced changes in NS observables. As both types of perturbation make the EoS softer, they lead to a smaller mass, smaller radius, and greater tidal deformability for a given NS with fixed central pressure. 
Note that while a Gauss function perturbation only changes the $M$-$R$ and $\Lambda$-$R$ curves within a finite range of $P_c$, a step function perturbation changes the whole curve with $P_c$ above the PT pressure.

\begin{figure*}[!hbtp]\centering
    \includegraphics[width=0.95\textwidth]{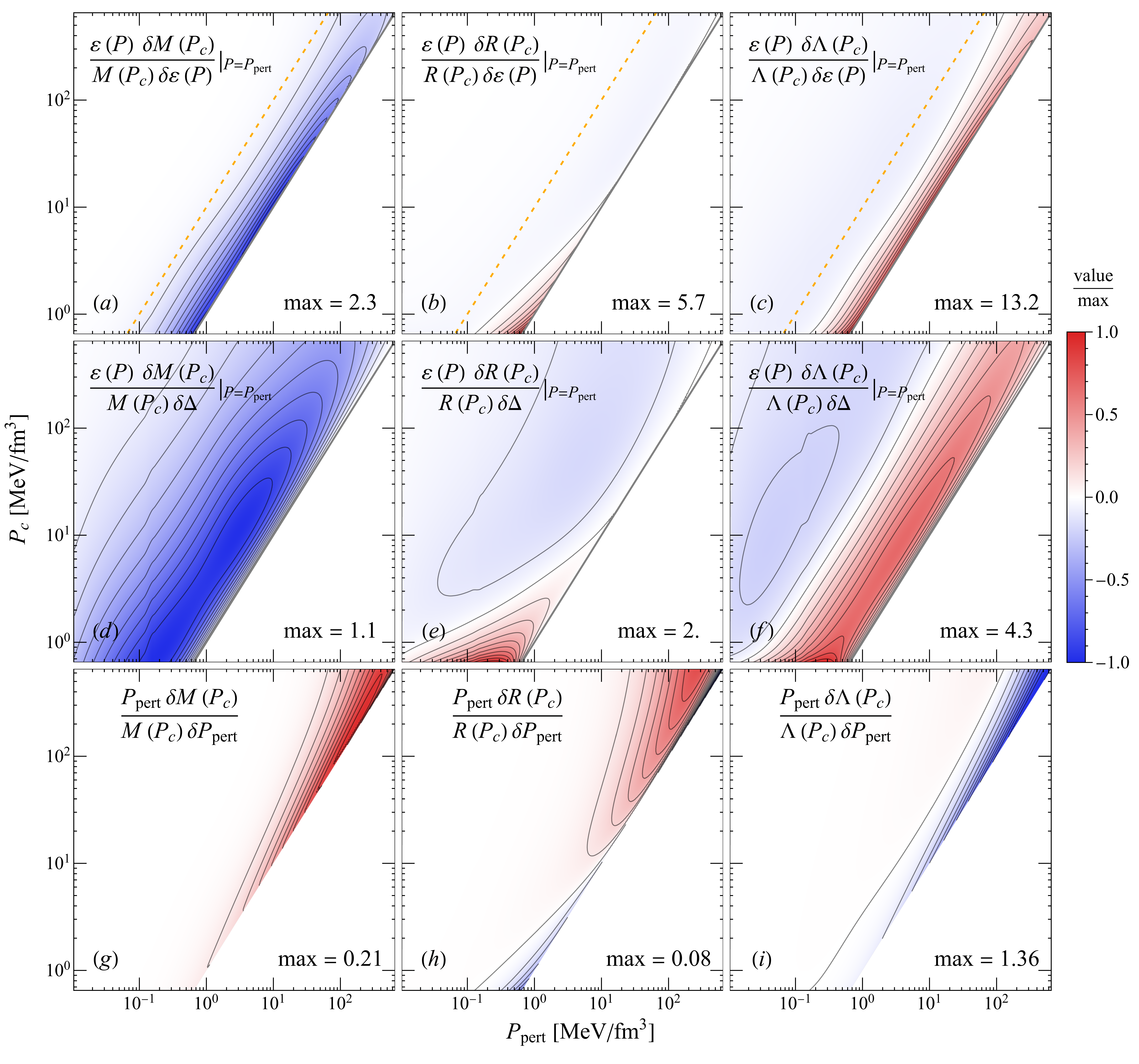}
    \caption{Sensitivity of the NS mass (left), radius (middle), and tidal deformability (right) with respect to a local change [top, c.f. \protect{Eq.~\eqref{eq.eos_sum_perturbation}}] or the latent heat (mid) and transition pressure (lower) of a first-order phase transition [c.f. \protect{Eq.~\eqref{eq:pert_PT}}] on top of the \texttt{SFHo} equation of state. In the upper panels, orange dotted lines are added to indicate that $P_\mathrm{pert} = P_c/10$. In each sub-figure, the color represents the sensitivity scaled by the maximum value, which is displayed in the lower-right corner.}
    \label{fig:sensitivity}
\end{figure*}

\subsubsection{Perturbation in the central pressure}
\label{sec:algorithm:linear:preussure}

To find the linear response against the change in the central pressure, we express $O \in \{v, m, y\}$ in an integration manner, $O(\xi) = \int_{\xi}^{\xi_c} \mathcal{K}_{O}\, \mathrm{d}\zeta\,, \,
$where $\xi_c\equiv\ln\frac{P_c}{P_\text{bnd}}$.
When changing the central pressure, $\xi_c \to \xi_c + \delta \xi_c$, we denote that $\lambda \equiv 1 + \frac{\delta \xi_c}{\xi_c}$, the perturbed quantities are given by
\begin{align}
\tilde{O}(\lambda \xi) =\;& 
   \int_{\lambda \xi}^{\lambda \xi_c} \mathcal{K}_{O}\,  \mathrm{d} \zeta =
   \int_{\xi}^{\xi_c} \mathcal{K}_{O}|_{P\to P_\text{bnd} e^{\lambda \zeta}}\,\lambda \mathrm{d} \zeta\,,
\end{align}
where we have performed the variable substitution $\zeta \to \lambda \zeta$ in the second equality.
Then, the variations of quantities read
\begin{align}
\begin{split}
&\delta O(\xi) \equiv
    \tilde{O}(\lambda \xi) - O(\xi)\\
=\;&
    \int_{\xi}^{\xi_c} 
    \bigg[
      \frac{\delta \xi_c}{\xi_c}\mathcal{K}_{O} 
    + \mathcal{K}_{O,y} \delta y
    + \mathcal{K}_{O,v} \delta v
    + \mathcal{K}_{O,m} \delta m
\\&
    + \Big(\mathcal{K}_{O,P} +
    \mathcal{K}_{O,\varepsilon} \kappa_s + \mathcal{K}_{O,\kappa_s} \frac{\mathrm{d}\kappa_s}{\mathrm{d}P}\Big)\frac{\zeta \, p\,\delta \xi_c}{\xi_c}\bigg] \mathrm{d}\zeta\,,
\end{split}
\end{align}
which can be computed according to the differential equation
\begin{align}
\begin{split}
0 =\;&
    \frac{\mathrm{d}}{\mathrm{d} \xi} \frac{\delta O}{\delta \xi_c}
    +  \mathcal{K}_{O,y} \frac{\delta y}{\delta \xi_c}
    +  \mathcal{K}_{O,v} \frac{\delta v}{\delta \xi_c}
    + \mathcal{K}_{O,m} \frac{\delta m}{\delta \xi_c}
\\+&
    \Big(\mathcal{K}_{O,P} +
    \mathcal{K}_{O,\varepsilon} \kappa_s + \mathcal{K}_{O,\kappa_s} \frac{\mathrm{d}\kappa_s}{\mathrm{d}P}\Big) \frac{P\,\xi}{\xi_c}
    + \frac{\mathcal{K}_{O}}{\xi_c}\,,
\end{split}
\label{eq.dX_dPc_diff}
\end{align}
with $\mathcal{K}_{m,y} = \mathcal{K}_{v,y} = \mathcal{K}_{m,\kappa_s} = \mathcal{K}_{v,\kappa_s} = 0$.
Given that different points on a $M$-$R$-$\Lambda$ curve are different in $P_c$, we note that the derivatives, $\frac{\partial M}{\partial P_c}$, $\frac{\partial R}{\partial P_c}$, and $\frac{\partial \Lambda}{\partial P_c}$, determine the slope of the curve.

\subsubsection{Variation analysis: summary}
\label{sec:variation:summary}

Here we summarize the variation analysis procedures. 
Given perturbations in the EoS function and in the central pressure, we solve Eqs.~(\ref{eq:eom:dv_de_diff}--\ref{eq:eom:dy_de_diff}, \ref{eq:eom:arb_gradient}) and \eqref{eq.dX_dPc_diff}, respectively, and obtain the derivatives at the $P=P_\mathrm{bnd}$ edge, 
\begin{align}
\frac{\partial M}{\partial \vartheta_k}
\equiv\;&
    \frac{\delta M}{\delta \vartheta_k}
=    
    \frac{\delta m(\xi)}{\delta \vartheta_k}\Big|_{\xi=0}\,,
    \label{eq.dev_m_P}\\
\frac{\partial R}{\partial \vartheta_k}
\equiv\;&
    \frac{\delta R}{\delta \vartheta_k}
=   
    \frac{\delta v(\xi)}{3R^2\delta \vartheta_k}\Big|_{\xi=0}\,,
    \label{eq.dev_r_P}\\
\frac{\partial Y}{\partial \vartheta_k}
\equiv\;&
    \frac{\delta Y}{\delta \vartheta_k}
=    
    \frac{\delta y(\xi)}{\delta \vartheta_k}\Big|_{\xi=0}\,,
    \label{eq.dev_y_P}
\end{align}
where $\bth$ is the shorthand of all the EoS parameters and the central pressures of different NSs.
A numerical validation of Eqs.~(\ref{eq:eom:dv_de_diff}--\ref{eq:eom:dy_de_diff}, \ref{eq:eom:arb_gradient}) and \eqref{eq.dX_dPc_diff} is provided in Appendix~\ref{app:validation}.
The response of $\Lambda$ with respect to a perturbation in $\vartheta_k$ can be given by
\begin{align}
\frac{\partial \Lambda}{\partial \vartheta_k}
=\;&
    \frac{\partial \Lambda}{\partial Y} 
    \frac{\partial Y}{\partial \vartheta_k}
    +
    \frac{1}{R}\frac{\partial \Lambda}{\partial \beta} 
    \Big(\frac{\partial M}{\partial \vartheta_k}
    -\beta \frac{\partial R}{\partial \vartheta_k}\Big)\,.
    \label{eq.dev_Lambda}
\end{align}

With the variations computed in an efficient manner, we are able to update the model parameters in the numerical reconstruction of the EoS. Moreover, the variational analysis is valuable by itself --- it directly gives the sensitivity of the NS mass, radius, and tidal deformability with respect to the change in the equation of state.
In Fig.~\ref{fig:sensitivity}, we start from the \texttt{SFHo}~\cite{Steiner:2012rk} equation of state, $\varepsilon = \varepsilon_\mathrm{SFHo}(P)$, and respectively introduce two types of perturbation in the energy density. 

First, we add a small Dirac-$\delta$ function local perturbation in the EoS, $\varepsilon(P) \to \varepsilon(P) + \delta(P - P_\mathrm{pert})\, \delta_\varepsilon / P_\mathrm{pert}$, and compute the change in mass for a NS with its central pressure being $P_c$, denoted as $\delta_M(P_c)$, and the functional derivative can be obtained as $\frac{\delta M(P_c)}{\delta \varepsilon(P)}|_{P=P_\mathrm{pert}} = \frac{\delta_M(P_c)}{\delta_\varepsilon}$. 
Varying the position of local change ($P_\mathrm{pert}$) and the NS of interest ($P_c$), we present the dimensionless relative functional derivatives (scaled by the respective value of $M$ and $\varepsilon(P_\mathrm{pert})$) in  Fig.~\ref{fig:sensitivity}, panel ($a$), and 
likewise for $R$ and $\Lambda$ in panels ($b$) and ($c$). 

Second, we add a small perturbation in the transition pressure and latent heat of a first-order phase transition at $P = P_\mathrm{pert}$, 
\begin{align}
    \varepsilon(P) = \varepsilon_\mathrm{SFHo}(P) + \Theta(P-P_\mathrm{pert}) \Delta.
    \label{eq:pert_PT}
\end{align}
For a NS with central pressure $P_c$, we compute the change of $M$, $R$, and $\Lambda$, and show their scaled, relative deviations in Fig.~\ref{fig:sensitivity}, panels ($d$-$i$).

Let us take a NS with $P_c = 10~ \text{MeV}/\text{fm}^3$ for example: a phase transition at $P_\mathrm{pert} = 1~\text{MeV}/\text{fm}^3$ corresponds to $\frac{\varepsilon(P) \delta M(P_c)}{M(P_c)\delta \Delta} \approx -0.77$, meaning that a phase transition at $P=1~\text{MeV}/\text{fm}^3$ with latent heat $\Delta = r\times \varepsilon(P=1~\text{MeV}/\text{fm}^3)$ would lead to a change in mass as $M \to M - 0.77r\times M$.
In each sub-figure, we measure the maximum of the absolute value for the shown region and display it in the lower right corner. 
We identify the systematic trend that the scaled derivatives of $\Lambda$ always have the greatest values, whereas those for $M$ are always the smallest, which implies that the tidal deformability is more sensitive to the change in the EoS. 

For a local change in the EoS, we note that the scaled derivatives are approaching zero for $P_\mathrm{pert} \lesssim P_c/10$, i.e., to the left of the orange dotted lines. This indicates that the mass, radius, and tidal deformability of a NS are not sensitive to any local changes in the EoS at pressures smaller than $\sim 1/10$ of its central pressure. 
Consequently, NS observables are  sensitive to the EoS for a range of pressures $P \in (P_c/10, P_c]$, and one is not able to infer the EoS for pressures $\lesssim 1/10$ of the relevant pressure in NSs. Other prior physics knowledge must be required in a global inference for the low-pressure region.
In the lower panels, the scaled derivatives are not vanishing since a first-order phase transition~\eqref{eq:pert_PT} introduced at $P=P_\text{pert}$ would also influence the EoS at the adjacency of $P=P_c$. 

Finally, note that the maximum values in ($g$), ($h$) are significantly smaller than those in ($a$), ($b$), ($d$), and ($e$), indicating that masses and radii are less sensitive to the pressure where PT happens compared to other changes in the EoS.

\subsection{Parametrization of the Equation of State}
\label{sec:algorithm:parameterization}

In the BA of the EoS, it is important to implement physical constraints, for instance, the causality requirement $\kappa_s \equiv c_s^{-2} = \frac{\mathrm{d}\varepsilon}{\mathrm{d}P} > 1$.
Thus, we represent the EoS by the inverse speed of sound squared as a function of the log-pressure, $\kappa_s(\xi)$, and always require it to be greater than unity. We assume the low temperature limit for NS EoS as $T \ll \mu$ in cold dense matter,
and $\varepsilon = -P + \mu\, n$ with $\mu=\mathrm{d}\varepsilon/\mathrm{d}n$ and $n=\mathrm{d}P/\mathrm{d}\mu$. 
Therefore, 
\begin{align}
    \frac{\mathrm{d}\varepsilon}{\varepsilon+P} = \frac{\mathrm{d}n}{n}\,,
\end{align}
and henceforth
\begin{align}
    n(\xi) = n_0\, \exp\bigg(\int_{\xi_0}^{\xi}\frac{\kappa_s(\zeta)}{1 + \frac{\varepsilon(\zeta)}{P(\zeta)}} \mathrm{d}\zeta \bigg),
\end{align}
with $P(\xi)\equiv P_\text{bnd} e^{\xi}$ and the boundary conditions are set to be $P_\text{bnd} = 1.4 \times 10^{-12}~\mathrm{MeV/fm}^{3}$, $P_{0} = 10^6 P_\text{bnd}$, and $n_0\equiv n(P_{0}) = 2.4 \times 10^{-6}~\mathrm{fm}^{-3}$ does not rely on models\footnote{We extracted the value from the \texttt{SFHo} EoS~\cite{Steiner:2012rk}. Also, note that $n_0$ here is \textit{not} the nuclear saturation density ($\approx 0.16~\mathrm{fm}^{-3}$).}.

We parametrize the inverse speed of sound squared containing both regular parts and Dirac-$\delta$ functions corresponding to possible first-order phase transitions,
\begin{align}
\kappa_s(\xi) =\;&
    \mf(\xi|\boldsymbol{\theta}) + \sum_l \frac{\Delta_l}{P(\xi_l)} \delta(\xi-\xi_l)\,,
\\
\begin{split}
\varepsilon(\xi) =\;&
    \varepsilon(0)
    +\int_{0}^{\xi} \mf(\zeta|\boldsymbol{\theta})
    P(\zeta) \mathrm{d}\zeta
\\&
    +\sum_l \Delta_l \Theta(\xi-\xi_l)\,,
\end{split}\\
    \mf(\xi|\boldsymbol{\theta}) =\;& \sum_l \Theta(\xi -\xi_{l-1})\Theta(\xi_{l} -\xi) \mf_{l}(\xi|\boldsymbol{\theta}_l),
\end{align}
where $\Theta$ is the step function, $\boldsymbol{\theta}$ are parameters of the regular sector, and $\xi_l$ indicates the log-pressure of the first-order phase transition point, and $\Delta_l \geq 0$ the latent heat. 
Keeping in mind that there could be rich phase structures, we have introduced a parametrization scheme that allows alternative number of PT points, and the number of $l$'s is a hyperparameter of the parametrization.
$\mf$ is represented as piece-wise functions accounting for the fact that the speed of sound value could be discontinuous at the phase transition point.
Each $\mf_{l}$ is parametrized as $\mf_{l}(\xi|\boldsymbol{\theta}_l) = 1+e^{{f}_{l}(\xi|\boldsymbol{\theta}_l)}$ to fulfill the causality condition.

For computing the $\boldsymbol{\theta}$-derivatives of $\chi^2$ and possible prior constraints on the EoS, it would be useful to first write down the parameter derivatives for the energy density,
\begin{align}
\begin{split}
\frac{\partial \varepsilon(\xi)}{\partial \theta_k}
=\;&
    \int_{0}^{\xi} \frac{\partial\mf(\zeta|\boldsymbol{\theta})}{\partial \theta_k}
    P(\zeta) \mathrm{d}\zeta,\\
\frac{\partial \varepsilon(\xi)}{\partial \Delta_l}
=\;&
    \Theta(\xi-\xi_l)\,,\\
\frac{\partial \varepsilon(\xi)}{\partial \xi_l}
=\;&
    \big(\mf_l(\xi_l)-\mf_{l+1}(\xi_l)\big)
    P_l\,\Theta(\xi-\xi_l)
\\&
    -\Delta_l \delta(\xi-\xi_l)\,,
\end{split}
\label{eq.dev_e}
\end{align}
and then for the number density,
\begin{align}
\begin{split}
\frac{\partial n(\xi)}{n\,\partial \theta_k}
=\;&
    \int_{0}^{\xi} \Bigg(
    \frac{\partial_{\theta_k}\mf(\zeta|\boldsymbol{\theta})}{1 + \frac{\varepsilon(\zeta)}{P(\zeta)}}
-
    \frac{\kappa_s(\zeta)\partial_{\theta_k} \varepsilon(\zeta)}{\big(1 + \frac{\varepsilon(\zeta)}{P(\zeta)}\big)^2P(\zeta)} \Bigg)\mathrm{d}\zeta\,,\\
\frac{\partial n(\xi)}{n\,\partial \Delta_l}
=\;&
    \Bigg(\frac{1}{\varepsilon_l + P_l}
    -
    \int_{\xi_l}^{\xi}
    \frac{\kappa_s(\zeta)\,\mathrm{d}\zeta}
    {\big(1 + \frac{\varepsilon(\zeta)}{P(\zeta)}\big)^2P(\zeta)}\Bigg) 
    \Theta(\xi-\xi_l)
    \,,\\
\frac{\partial n(\xi)}{n\,\partial \xi_l}
=\;&
    -\frac{\Delta_l}{\varepsilon_l + P_l} \Big(\delta(\xi-\xi_l) +\frac{\Theta(\xi-\xi_l)}{1+\varepsilon_l/P_l}\Big) 
\\&
    +\Big(\mf_l(\xi_l)-\mf_{l+1}(\xi_l)\Big)\times
\\&
    \quad
    \bigg(\frac{\Theta(\xi-\xi_l)}{1+\varepsilon_l/P_l}
    -P_l\,\int_{\xi_l}^{\xi}
    \frac{\kappa_s( \zeta ) P( \zeta ) \mathrm{d}\zeta}{(\varepsilon( \zeta ) + P( \zeta ) )^2}\bigg)\,,
\end{split}
\label{eq.dev_n}
\end{align}
where we introduced the shorthands that $P_l\equiv P(\xi_l)$ and $\varepsilon_l \equiv \varepsilon(\xi_l+0^+)$.

Furthermore, for NS variables $O \in \{v(\xi), m(\xi), y(\xi) \}$ at $\xi=0$, their boundary values used in Eqs.~(\ref{eq.dev_m_P}--\ref{eq.dev_y_P}) can be obtained 
\begin{align}
\begin{split}
    \frac{\delta O}{\delta \theta_k} =\;& 
    \int_{0}^{\xi_c}\mathrm{d}\zeta\, \frac{\partial \mf
    (\zeta|\boldsymbol{\theta})}{\partial \theta_k}
    P(\zeta)
    \int_{\zeta}^{\xi_c}\mathrm{d}\eta\, \Delta_O(0|\eta),
\end{split}
\\
    \frac{\delta O}{\delta \Delta_l} =\;& \int_{\xi_l}^{\xi_c}\mathrm{d}\zeta\,\Delta_O(0|\zeta) ,
\\
\begin{split}
    \frac{\delta O}{\delta \xi_l}  =\;&
    \big(\mf_l(\xi_l)-\mf_{l+1}(\xi_l)\big) P_l
    \int_{\xi_l}^{\xi_c}\mathrm{d}\zeta\,\Delta_O(0|\zeta)
\\&    
    - \Delta_l \Delta_O(0|\xi_l).     
\end{split}
\end{align}
As a reminder, the perturbative responses of NS variables, $\Delta_O(\xi|\xi')$, are obtained by solving Eqs.~(\ref{eq:eom:dv_de_diff}--\ref{eq:eom:dy_de_diff}).
Finally, the parameter gradient of the $\chi^2$-function can be obtained  
\begin{align}
\frac{\nabla_{\bth} \chi^2}{2} =\;& \sum_{i} \sum_{O\in\{M,R,\Lambda\}}
    \frac{\partial \chi_i^2(\bth)}{2\partial O_i} \nabla_{\bth} O_{i}\,,
\end{align}
which can be further simplified for uncorrelated Gaussian distributions~\eqref{eq:chi_sq_Gaussian},
\begin{align}
\frac{\nabla_{\bth} \chi^2}{2} =\;& \sum_{i} \sum_{O\in\{M,R,\Lambda\}}
    \frac{O_{i} - \overline{\text{O}}_{i}}{\Delta_{O,i}^2} \nabla_{\bth} O_{i}\,.
\label{eq.grad_chi}
\end{align}

Here we outline the procedures of finding 
$\bth_\mathrm{opt}$ according to the gradient-based iteration method:
\begin{itemize}
\item[(a)] 
    Start from an arbitrary parameter set, $\bth^{(0)}$, and then repeat the iterations (b) and (c).
\item[(b)] 
    At the $n^\mathrm{th}$ iteration with parameter set $\bth^{(n)}$, compute the EoS, solve the TOV equations and the differential equations for the variations~(\ref{eq:eom:dv_de_diff}--\ref{eq:eom:dy_de_diff}, \ref{eq.dX_dPc_diff}), and compute $\nabla_{\bth}\chi^2|_{\bth = \bth^{(n)}}$ according to \eqref{eq.grad_chi}.
\item[(c)] 
    Update the parameter set $\bth^{(n+1)} = \bth^{(n)} + \alpha \,\boldsymbol{\Delta}_{\bth}$, where $\alpha$ is referred to as the learning rate, and the update step $\boldsymbol{\Delta}_{\bth}$ is determined by $\nabla_{\bth}\mathcal{J}$.
    For first-order gradient descent, $\boldsymbol{\Delta}_{\bth} = -\nabla_{\bth}\mathcal{J} $. In this work, we used the Adaptive Moment Estimation (also known as ADAM)~\cite{2014arXiv1412.6980K, Mehta:2018dln} algorithm which includes a momentum term to avoid saturation at the local minimum of $\mathcal{J}$ and takes into account the second order correction to accelerate the learning procedure. 
\item[(d)]
    Repeat (b) and (c) until $\boldsymbol{\Delta}_{\bth}$ is small enough and the loss function becomes stable. Now, $\bth$ is optimized to the value that minimizes the loss function.
\end{itemize}

Once the most optimal point, $\bth^\mathrm{opt}$, is obtained, one can perform BA by invoking importance sampling~\cite{Kahn:1953}.
We first compute a reference distribution according to the first-order approximation for the Posterior distribution --- a correlated Gaussian distribution centered at the optimal point
\begin{align}
    \mathrm{P}_\text{ref}(\bth) = \mathcal{N}_0 \, e^{- \frac{1}{2}C_{kk'}(\vartheta_k-\vartheta^{\mathrm{opt}}_{k})(\vartheta_{k'}-\vartheta^{\mathrm{opt}}_{k'})},
     \label{eq.reference}
\end{align}
where the covariance matrix is given by the parameter gradients (with $O_i, O_i' \in \{M_i, R_i\}$)
\begin{align}
\begin{split}
{C}_{kk'} =\;& 
    \frac{1}{2}\frac{\partial^2 \mathcal{J}}{\partial\vartheta_k\, \partial\vartheta_{k'}}
\\\approx\;&
    \frac{1}{2}\frac{\partial^2 \mathcal{R}}{\partial\vartheta_k\, \partial\vartheta_{k'}}
    + \frac{1}{4} \sum_{i,O_i,O_i'}\frac{\partial^2 \chi^2}{\partial O_i\, \partial O_i'} \frac{\partial O_i}{\partial\vartheta_k} \frac{\partial O_i'}{\partial\vartheta_{k'}}\,.
\end{split}\label{eq:correlation_kernal}
\end{align}
With the reference distribution, we take the following procedures to obtain the distribution of EoSs:
\begin{itemize}
    \item[(a)] Sample a parameter set, $\bth_n$, according to the reference distribution~\eqref{eq.reference};
    \item[(b)] Compute the corresponding EoS $\varepsilon_n = \varepsilon(P|\bth_n)$;
    \item[(c)] Solve the TOV equations and compute the loss function $\mathcal{J}(\bth_n)$;
    \item[(d)] Collect $\varepsilon_n$ into a histogram, with a weight $w_n \equiv \frac{e^{-\mathcal{J}(\bth_n)}}{\mathrm{P}_\text{ref}(\bth)}$.
\end{itemize}
By repeating (a)-(d) and collecting sufficient numbers of samples, we obtain the posterior distribution of EoS. 
It should be noted that the true $\mathrm{P}(\boldsymbol{\vartheta})$ is not necessarily Gaussian, and one may still apply this framework to extract the EoS distribution from realistic observations of NS masses and radii, which follow nonGaussian distribution. 
It is worth noting that we parametrize the EoS by always constraining the speed of sound to be positive and causal, so that any EoS sampled from the final posterior distribution is physical. 
We supplement in Appendix~\ref{app:importance_sampling} for a proof of the importance sampling method and its uncertainty estimation.

\vspace{3mm}
\emph{Parameterization with DNN}. --- While what have been discussed above (including general discussions of the algorithm in Sec.~\ref{sec:algorithm:general} and~\ref{sec:algorithm:linear}) are applicable for arbitrary parametrization of the EoS, we exploit the deep neural network (DNN) as an unbiased and flexible parametrization~\cite{Leshno93multilayerfeedforward, Kratsios_2021} of the EoS.

DNN is a parametrization scheme which can approximately express any $\mathbb{R}_n \to \mathbb{R}_m$ function mapping between independent variables (\textit{inputs}) $\boldsymbol{x} = \{x_1, \cdots, x_n\}$ and dependent variables (\textit{outputs}) $\boldsymbol{y}=\{y_1,\cdots,y_m\}$, $\boldsymbol{y}=\boldsymbol{y}(\boldsymbol{x})$. 
It constructs the functional form by iteratively composing $N$ simple building blocks (also called \textit{layer} representing a vector-to-vector function). 
Each layer performs a linear transformation on the output from the preceding layer, followed by an element-wise nonlinear transformation dictated by the \textit{activation function} $\sigma^{(\ell)}(z)$ 
\begin{equation}
a^{(\ell)}_s = \sigma^{(l)}(z^{(\ell)}_{s}),\qquad z^{(\ell)}_{s} \equiv {b}^{(\ell)}_{s} + \sum_{t} {W}^{(\ell)}_{st} {a}^{(\ell-1)}_{t}\,, \label{eq:dnn:iteration}
\end{equation}
for $s = 1, \cdots, n^{(\ell)}$ and $\ell = 1, \cdots, N$. 
The iteration starts from input variables, $a_s^{(0)} \equiv {x_s}$, and ends with the model output, $\boldsymbol{y}({\boldsymbol{x}}|\{W_{st}^{(\ell)},b_s^{(\ell)}\}) = \boldsymbol{a}^{(N)}$. At each layer, the activation function $\sigma(z)$ is usually chosen to be approximate functions of either a Heaviside step function or its integral, and each iteration~\eqref{eq:dnn:iteration} provides a piecewise zeroth-order or first-order interpolation of the target function. 
This compositional way of parametrization renders DNN an universal function approximator being able to fit any continuous function to arbitrary accuracy given enough hidden units~\cite{LeCun:2015pmr}. 

Along these lines, DNN delivers a powerful tool to learn about essential physical quantities in nuclear physics by capturing complex, nonlinear patterns that traditional models struggle with~\cite{Aarts:2025gyp}. For instance, DNN helps phase transition identification in heavy-ion collisions~\cite{Pang:2016vdc, OmanaKuttan:2020btb, Jiang:2021gsw} as well as in lattice study~\cite{Zhou:2018ill, Apte:2024vwn}, improves parton distribution function (PDF)~\cite{Karpie:2019eiq, NNPDF:2021uiq, Gao:2022iex} and spectral function reconstruction~\cite{Kades:2019wtd, Wang:2021jou, Shi:2022yqw}, quasi-particle modeling~\cite{Li:2022ozl, Li:2025csc}, extraction of heavy quark potentials~\cite{Shi:2021qri} and hadron-hadron interactions~\cite{Wang:2024ykk, Wang:2024dzc} from lattice QCD data, etc. It could also help nuclear many-body calculations as an unbiased but flexible \textit{Ansatz}~\cite{Adams:2020aax, Fore:2022ljl, Wang:2024xcs}. In addition, DNN-based emitting source functions offer new insights into particle production in heavy-ion collisions~\cite{Wang:2024bpl}.

In the present work, we use an $\mathbb{R}_1 \to \mathbb{R}_1$ network for each phase of nuclear matter, and each of them takes the log-pressure ($\xi$) as the input and the sound speed squared ($\mf_l$) as the output.
$W^{(\ell)}_{st}$ and $b^{(\ell)}_s$, respectively called \textit{weights} and \textit{biases} in DNN, are the model parameters ($\boldsymbol{\theta}$). 
Each DNN has $N=3$ layers, with widths of the intermediate layers (also known as \textit{hidden layers}) being $n^{(1)} = n^{(2)} = 128$. With the softplus activation function $\sigma^{(1)}(z) = \sigma^{(2)}(z) = \ln(1+e^z)$ for the hidden layers, we design a new activation function $\sigma^{(3)}(z) = 1+e^{z}$ to ensure causality requirement $\kappa_s>1$. 
At the $\ell^\mathrm{th}$ layer, there are $n^{(\ell)}\times n^{(\ell-1)}$ weights and $n^{(\ell)}$ biases, and therefore each $\mf_l(\xi|\boldsymbol{\theta})$ has $(128+128^2+128)+(128+128+1) = 16897$ parameters.

\begin{figure}[!hbtp]
    \centering
    \includegraphics[width=0.48\textwidth]{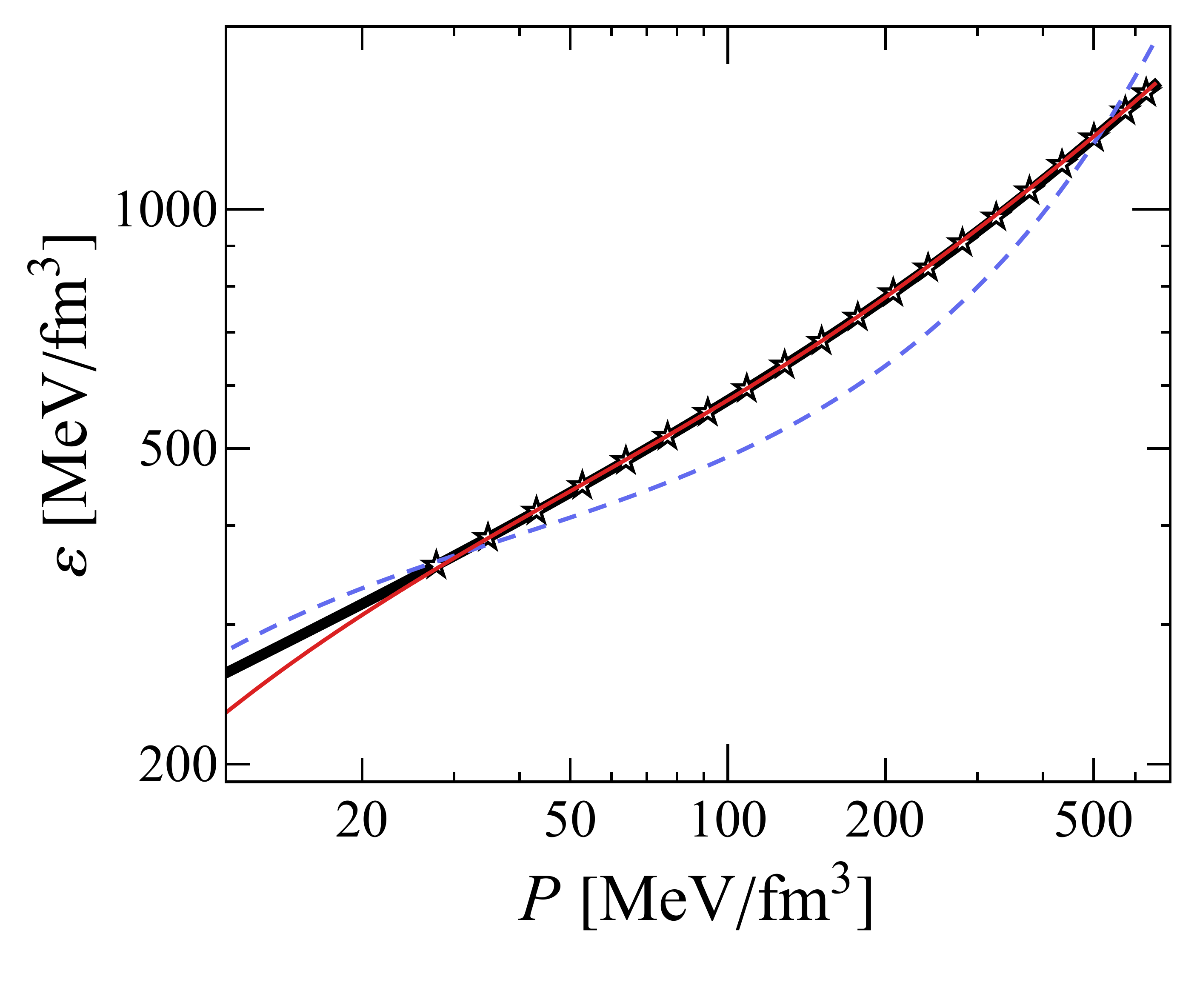}
    \includegraphics[width=0.48\textwidth]{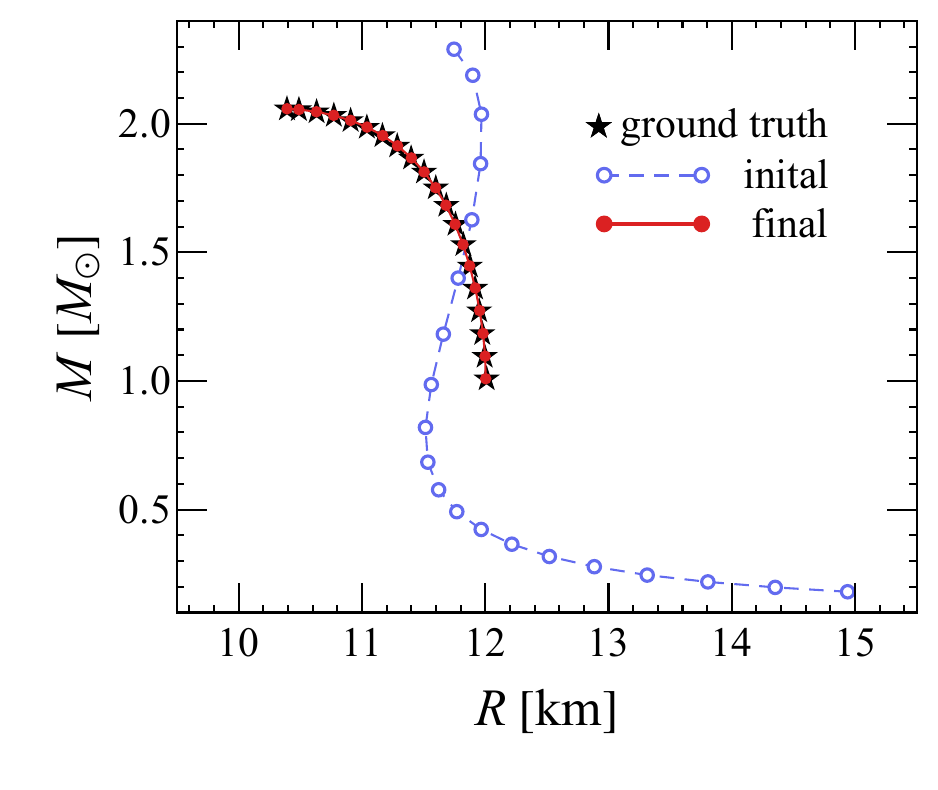}
    \caption{Reconstruction of Maximum a Posteriori EoS (upper) and the corresponding $M$-$R$ curve (lower) with a smooth known EoS, represented by the black thick line in the upper panel. Each black star in the lower (upper) panel represents the mass and radius (central pressure and energy density) of the a NS provided in the reconstruction. Blue dashed curves and open circles correspond to the initial condition of the network, while red solid curved and filled circles are for the final state.}
    \label{fig:map_smooth}
\end{figure}

\section{Numerical Reconstruction of EoS}
\label{sec:results}

With the computation tool set up, we are ready to check the quality of EoS reconstruction based on finite number of observations, especially the ability of identifying possible phase transitions. We will focus on artificial data of NSs' masses and radii given by known input of the EoS, with or without first-order PTs.
Such practices are also called \textit{mock test} in machine learning terminologies. Besides, we do not assume additional prior knowledge of the EoS except for causality which is already encoded in the neural network parametrization of the equation of state (NNEoS), that is, we take a flat Prior distribution for all the parameters ($\bth$).

\subsection{Maximum a Posteriori reconstruction of a smooth EoS}
\label{sec:results:map_smooth}

As a first step, we take an idealized limit and check the EoS reconstruction  from a set of NSs with wide coverage in the $M$-$R$ plot and with sufficiently small uncertainties. We take the \texttt{SFHo} EoS and compute twenty $M$-$R$ points with masses between one solar mass ($1\,M_{\odot}$) and the NS maximum mass, represented by black stars in Fig.~\ref{fig:map_smooth}. Uncertainties are set to be negligibly small and identical for all ``observational values'', and we keep the relative uncertainties to match, roughly, those of current observational data~\cite{Miller:2019cac, Riley:2019yda, Riley:2021pdl, Miller:2021qha, Reardon:2015kba, Gonzalez-Caniulef:2019wzi}, $\frac{\Delta_{M,i}}{\Delta_{R,i}} = \frac{M_{\odot}}{10 \, \text{km}}$. With the small uncertainty in masses and radii, we focus on the most optimal EoS that maximizes the posterior distribution, i.e., the Maximum a Posteriori EoS.

Regarding the reconstruction, we start from an NNEoS with initial condition being close to a realistic EoS but sizably different from SFHo (see the blue dashed curve with open circles in Fig.~\ref{fig:map_smooth}, lower panel), and then implement the optimization scheme to match the selected $M$-$R$ points. After sufficient iterations of parameter optimization, we reach a final state (red filled circles, lower panel) that all $M$-$R$ points are on top of their desired positions. The reconstructed equation of state (red curve, upper panel) is in good consistent with the \texttt{SFHo} ground truth (black), except for the low pressure regime where $P \,\lesssim 30 \mathrm{MeV}/\mathrm{fm}^3$, which corresponds to the lowest central pressure of all the artificially selected NSs. 
Observables of massive NSs (with high central pressures) 
are insensitive to the low-density EoS (see Fig.~\ref{fig:sensitivity}), and those of light-mass NSs allow degeneracy that changes in the EoS may compensate with each other so that the overall $M$-$R$ relation does not change. 

This exercise not only verifies the feasibility of using our optimization scheme and the NNEoS parametrization to reconstruct the nuclear matter equation of state, but also points out the region of reliability in the reconstructed EoS: one can only confidently reconstruct the EoS within the pressure range covered by the central pressures of the NSs; for other regions, additional prior physics knowledge would be needed to break the degeneracy. 
A thorough uncertainty estimation over the reconstructed EoS would also reflect the region of reliability, to which we devote for a future systematic study.

\begin{figure}[!hbtp]
    \centering
    \includegraphics[width=0.48\textwidth]{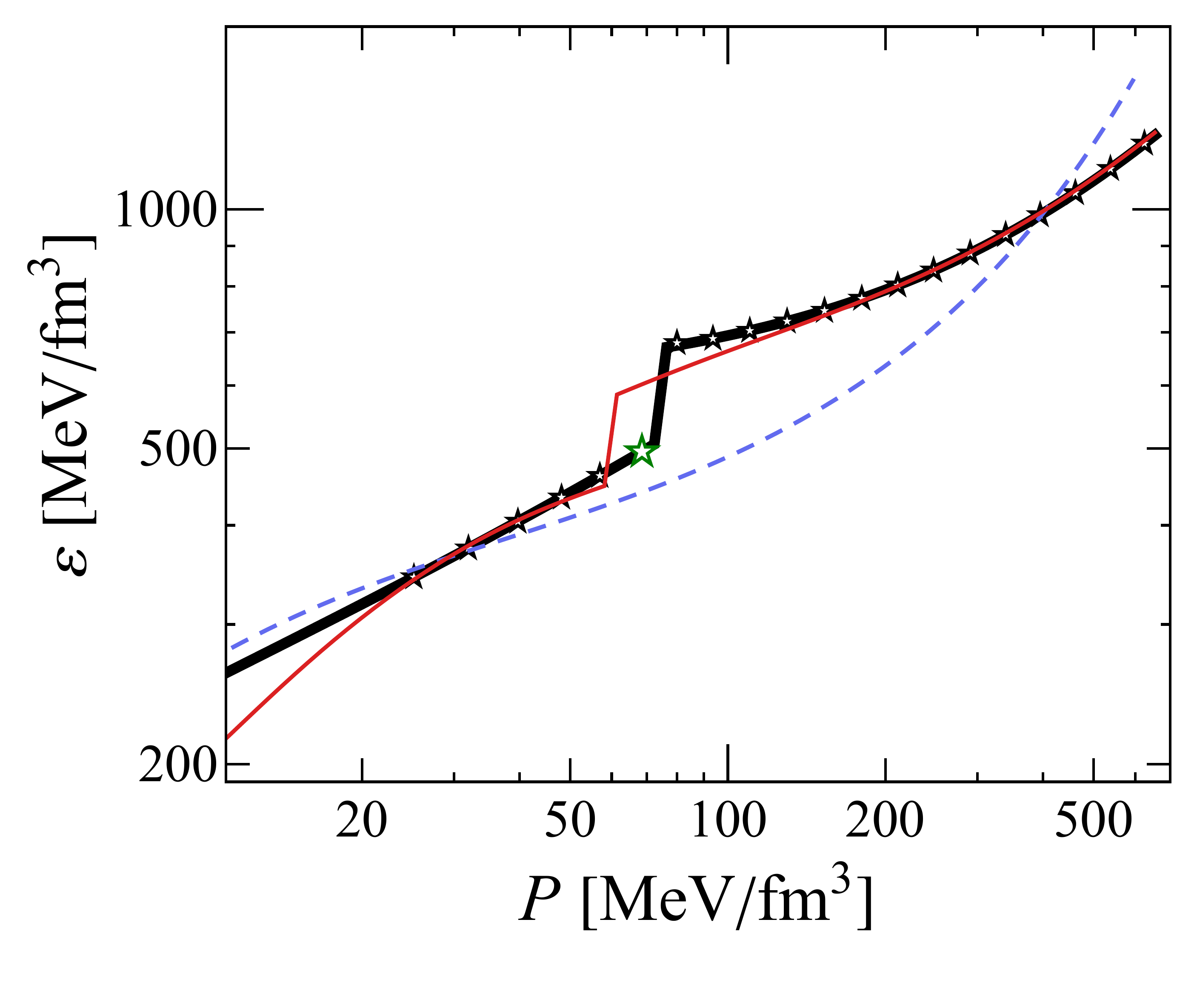}
    \includegraphics[width=0.48\textwidth]{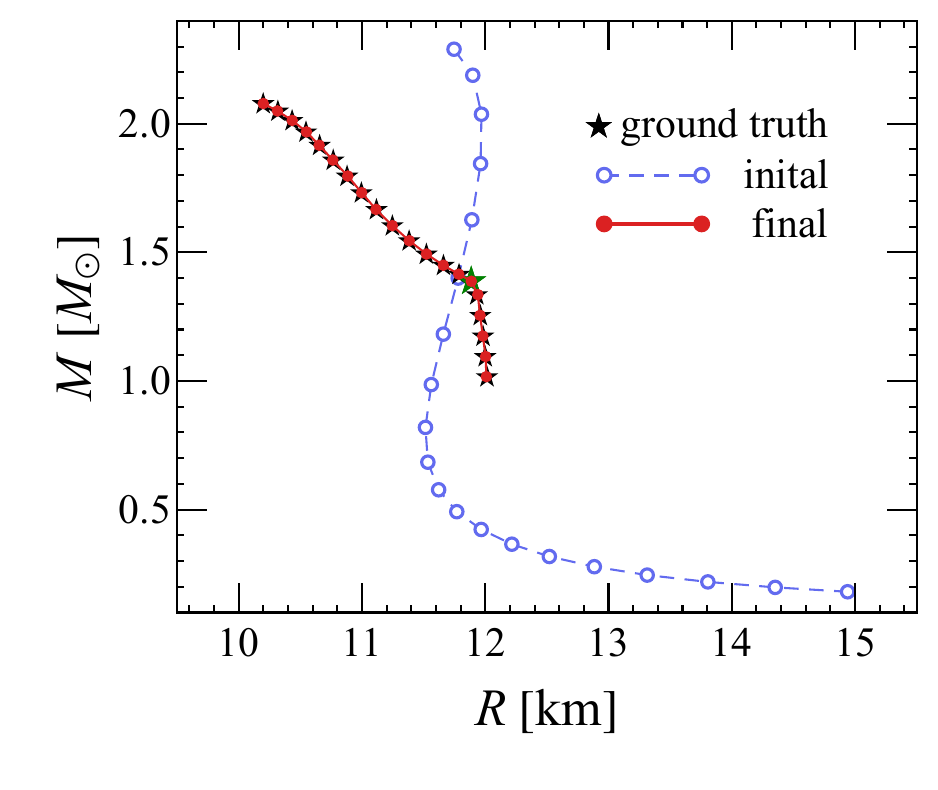}
    \caption{Same as Fig.~\protect{\ref{fig:map_smooth}} but for reconstructing EoS with a first-order phase transition. The green star indicates the NS with central pressure being closest to the phase transition pressure.}
\label{fig:map_withPT}
\end{figure}

\subsection{Maximum a Posteriori reconstruction of an EoS with first-order phase transition}
\label{sec:results:map_withPT}

Reconstructing NS EoSs that are smooth with machine learning techniques, indeed, has already been a relatively mature endeavor as demonstrated in previous studies, see e.g., Refs.~\cite{Soma:2022qnv, Soma:2022vbb} which were accomplished by some of the authors of the current work using a similar method. 
Rather than solving the TOV equations and computing their derivatives as in Sec.~\ref{sec:algorithm:linear}, Refs.~\cite{Soma:2022qnv, Soma:2022vbb} approximate the TOV equations by a TOV-solver network and exploit auto-differentiation to optimize the EoS. In addition to the aforementioned subtlety in estimating the systematic uncertainties associated with the TOV-solver-network approximation, the previously applied method exhibited difficulty in reconstructed data associated with first-order phase transitions, which is nevertheless of high interest to the community. 
The improved method devised in the present paper, in contrast, is suitable for reconstructing EoSs either with or without PTs. In this subsection, we follow the same procedure as in Sec.~\ref{sec:results:map_smooth} and examine the reconstruction quality for EoSs with first order PTs.

We adopt \texttt{SFHo} as the baseline EoS and introduce a PT with latent heat $\Delta\varepsilon = 150~\mathrm{MeV}/\mathrm{fm}^3$ at pressure $P_\mathrm{PT} =  76~\mathrm{MeV}/\mathrm{fm}^3$. Above the PT point, we take the stiffest (causal) limit that $c_s=1$. We employ twenty NSs marked by the star symbols in Fig.~\ref{fig:map_withPT} with the same uncertainty level as in the preceding subsection, and the reconstructed MAP NNEoS and its corresponding $M$-$R$ points are shown as the red curve (upper panel) and filled circles. 
It is evident that the reconstructed EoS agrees well with the ground truth for the regime covered by the NS's central pressures except for the adjacency of the PT point --- the reconstructed values for PT pressure and latent heat are, respectively, $P_\mathrm{PT} = 60~\mathrm{MeV}/\mathrm{fm}^3$ and $\Delta\varepsilon = 128~\mathrm{MeV}/\mathrm{fm}^3$. 

Despite of the sizable discrepancy between the reconstructed EoS and the ground truth, the reconstructed $M$-$R$ points are apparently well consistent with the mock observations, which indicates that the reconstructed EoS is also ``optimized'' given the finite set of $M$-$R$ points. From the mock observations, while one can confidently identify the discontinuity of the slope on the $M$-$R$ curve\footnote{Discontinuity in the slope of the $M$-$R$ curve is considered as signature of a first-order PT or substantial softening in this framework; see e.g. similar discussions in \cite{Chen:2019rja}.}, the discontinuity point itself is not obvious. 
To be specific, there could be many curves-with-one-fold that go through all the stars in Fig.~\ref{fig:map_withPT} (lower panel) and the fold should be adjacent to the point highlighted in green, but one can hardly tell whether it lies above or below. 
As a result, although the central pressure of the green star turns out to be below the ground truth of $P_\mathrm{PT}$, the reconstructed value of PT pressure falls below the former. This can also be seen from the sensitivity analysis shown in Fig.~\ref{fig:sensitivity}, which indicates that the observables of one NS is not sensitive to a PT right below its central pressure.

Based on this analysis, we conclude that even for the idealistic reconstruction with negligible uncertainties, finite number of NS observations could result in inaccurate reconstructed values of the PT pressure and, subsequently, the latent heat. Uncertainties in $P_\mathrm{PT}$ can be estimated from the central pressures of the NSs with $M$ and $R$ closest to the possible fold. 

\subsection{Marginal posterior distribution of phase transition parameters}
\label{sec:results:md}

\begin{figure}[htpb!]
\centering
\includegraphics[width=0.49\textwidth]{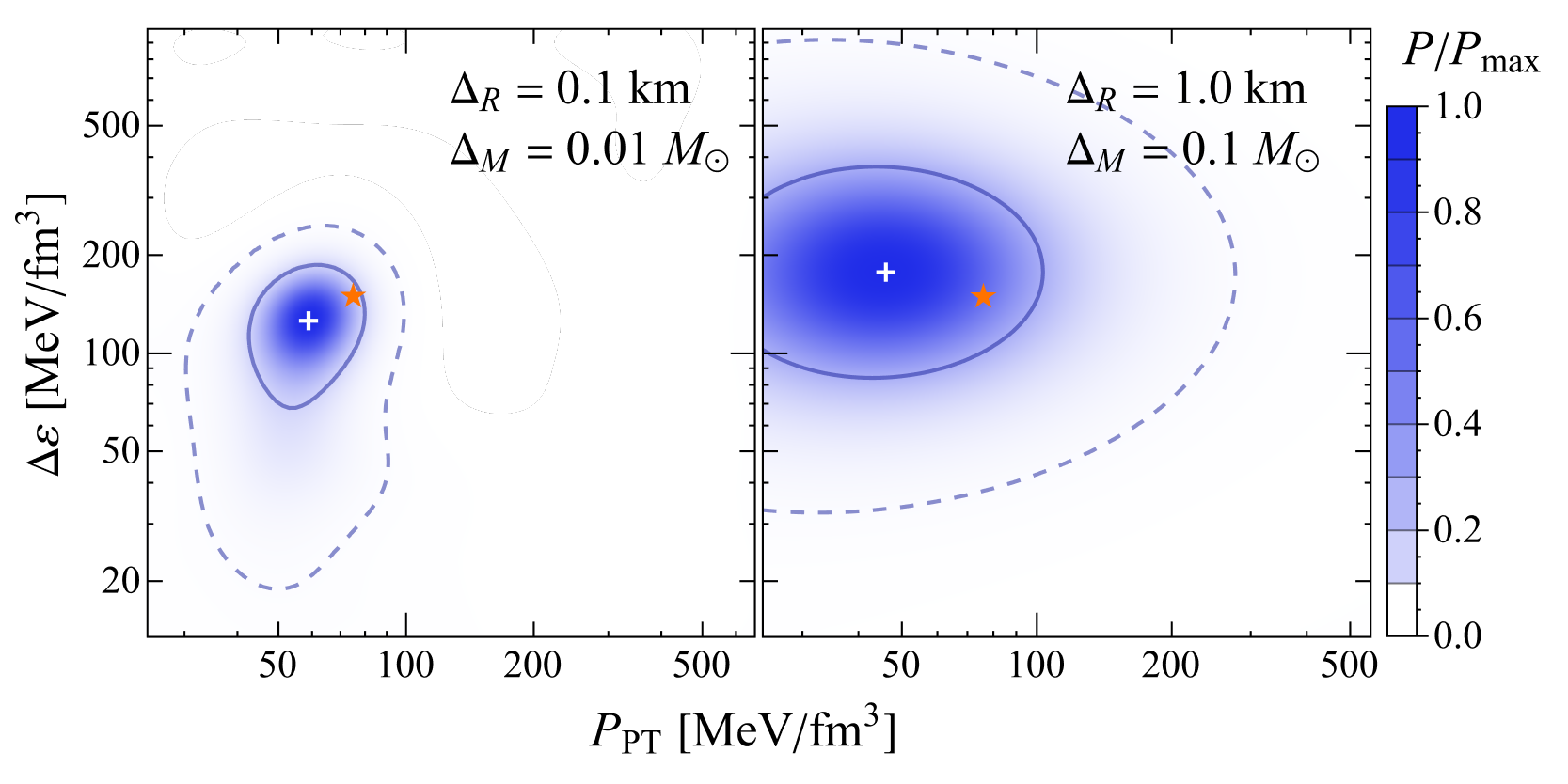}
\includegraphics[width=0.49\textwidth]{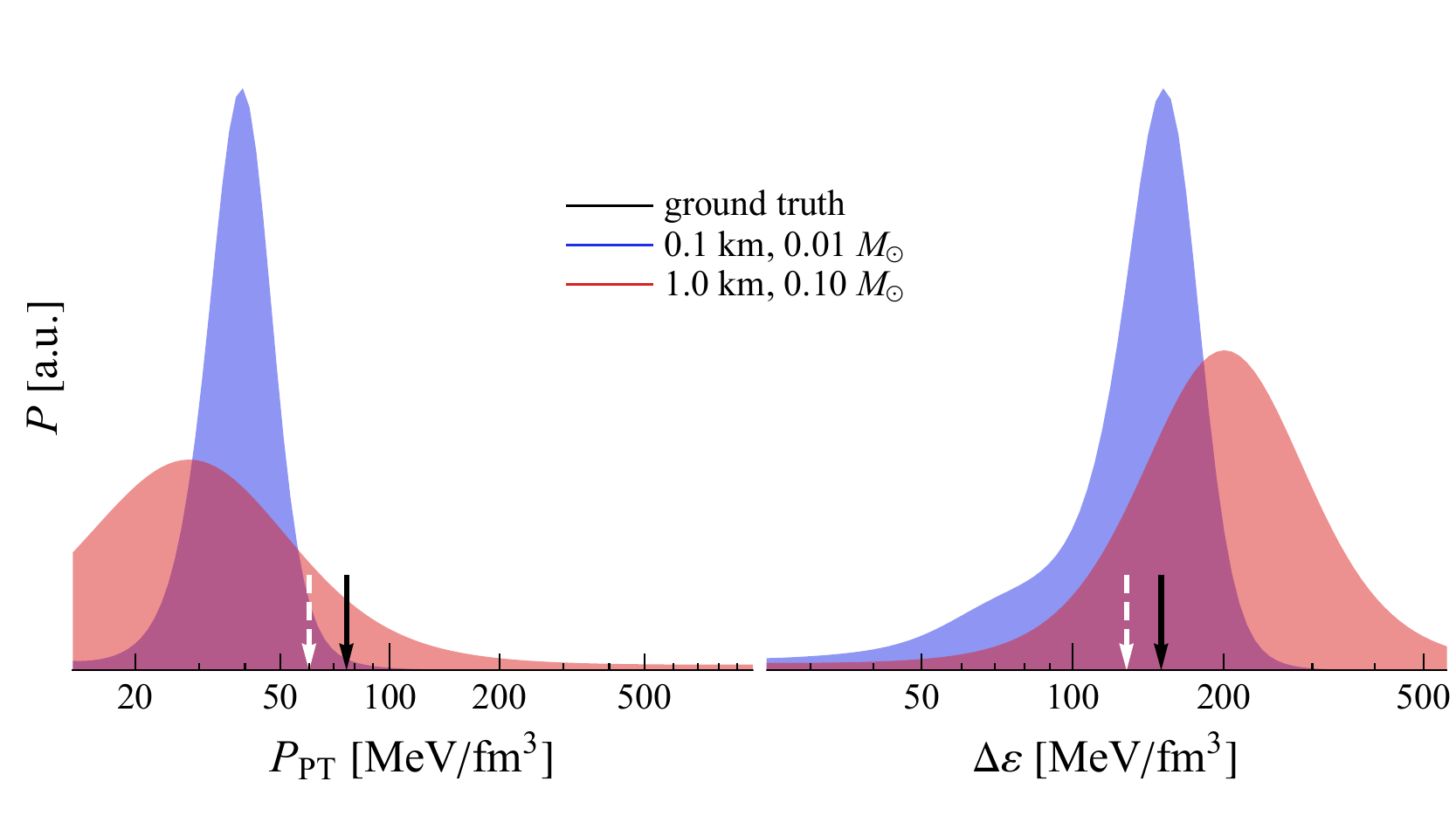}
\caption{Marginal posterior distributions of the first-order PT parameters, $P\,(\ln P_\mathrm{PT}, \ln \Delta\varepsilon)$ (upper), $P\,(\ln P_\mathrm{PT})$ (lower left), and $P\,(\ln \Delta\varepsilon)$ (lower right). 
Orange stars (black arrows) in the upper (lower) panel represent the ground-truth values $P_\mathrm{PT} =  76~\mathrm{MeV}/\mathrm{fm}^3$ and $\Delta\varepsilon = 150~\mathrm{MeV}/\mathrm{fm}^3$, and the white crosses indicate the values that maximize the marginal posterior. 
The mock observations with more optimistic (realistic) uncertainty levels is shown in the upper left (right) panel and blue (red) curves in the lower panels. 
In the upper panels, solid (dashed) curves indicate the $68\%$ ($95\%$) credible regions. In the lower panels, white dashed arrows represent the MAP values, $P_\mathrm{PT} =  60~\mathrm{MeV}/\mathrm{fm}^3$ and $\Delta\varepsilon = 128~\mathrm{MeV}/\mathrm{fm}^3$, as obtained in Sec.~\protect{\ref{sec:results:map_withPT}}.
    \label{fig:margin_posterior}}
\end{figure}

Noting the imperfectness in the reconstruction of first-order PT parameters even with precise but finite observations, it is important to correctly estimate the uncertainty of the EoS, which calls for a Bayesian analysis as illustrated in the description of algorithm. Being particularly interested in the uncertainties in the phase transition pressure and the associated latent heat, we focus on the marginal posterior distribution of $\xi_\mathrm{PT}$ and $\Delta\varepsilon$,
\begin{align}
    P(\xi_\mathrm{PT}, \Delta\varepsilon) \equiv \int P(\xi_\mathrm{PT}, \Delta\varepsilon, \boldsymbol{\theta}) \mathrm{d}\boldsymbol{\theta},
\end{align}
where $\boldsymbol{\theta}$ are NN parameters in the regular part of $\kappa_s$. 

With the detail of computation given in Appendix~\ref{app:importance_sampling:mpd}, we compute $P(\xi_\mathrm{PT}, \Delta\varepsilon)$ using the same setup as in the preceding subsection and consider two different levels of uncertainties: 
a more optimistic level with $\Delta_M = 0.01~M_{\odot}$, $\Delta_R = 0.1~\mathrm{km}$, and a more realistic level $\Delta_M = 0.1~M_{\odot}$, $\Delta_R = 1.0~\mathrm{km}$. 
Results are shown in Fig.~\ref{fig:margin_posterior}, where the orange stars represent the true PT parameters listed in the preceding subsection, whereas the white crosses indicate those maximizing the corresponding marginal posteriors. Note that they do not necessarily coincide with the MAP values listed above since other parameters have been integrated. 

It is evident that the true values are enclosed within the $68\%$ credible regions (CR), regardless of the uncertainty level being taken as optimistic or realistic. 
It is also natural that the optimistic case results in a narrower CR than the realistic one --- the better we constrain the masses and radii of the NSs, the more confident we are about the reconstructed PT parameters. 
Yet, the uncertainties of the reconstructed values do not linearly dependent on those of the NS observations: while both $\Delta_M$ and $\Delta_R$ in the optimistic case are one-tenth of those in the realistic case, the ratio of CR's in $P_\mathrm{PT}$ and $\Delta\varepsilon$ is greater than $1/10$. 
This is due to the fact that CR of the PT parameters not only depends on the precision of measurements, but also on the quantity of observed NSs with their central pressures being around the phase transition point. 
Both the number of NSs and the measure precisions determine the ``bottle neck'' of the PT reconstruction.

From the one-dimensional marginal posterior of the latent heat, we find that the $95\%$ CR is given by $\Delta\varepsilon=126^{+104}_{-79}~\mathrm{MeV}/\mathrm{fm}^3$ and $\Delta\varepsilon=193^{+399}_{-127}~\mathrm{MeV}/\mathrm{fm}^3$, respectively, 
at the optimistic and realistic uncertainty levels, from which one may estimate the ability of identifying a first-order PT: 
given the current precision level of measurements, one would be able to identify a strong first-order PT with latent heat of order $\Delta\varepsilon \sim 100~\mathrm{MeV}/\mathrm{fm}^3$, provided that the PT happens around $M \sim 1.4~M_{\odot}$ and a sufficient amount of $M$-$R$ observations taken place with a good coverage between $1~M_{\odot}$ and the maximum mass. 
The detectability  could be improved if (i) more precised measurements were achieved, and (ii) more NSs (in particular those around the PT critical mass) were observed.

\section{Summary and Discussions}
\label{sec:summary}

In this work, we performed a linear response analysis of the TOV equations and derived the analytical equations to calculate the derivatives of the NS observables with respect to the changes in the EoS and in the central pressure of the NS. 
Based on these investigations, we developed a computationally efficient algorithm to optimize an arbitrary parametrization of the EoS that best fits a finite set of the NS mass-radius observations. 
Bayesian posterior distributions of the EoS can also be computed for finite-precision measurements.

With the newly developed algorithm, we employed a general, unbiased parametrization of the EoS realized by neural networks, and demonstrated its ability in reconstruction of the EoS from NS measurements with and without a first-order PT. 
We further explored the ability of this method to reconstruct PT parameters from noisy mass and radius measurements. Based on the settings in the mock observations, we can make an order-of-magnitude estimate that measurements with the state-of-the-art precisions shall be able to reveal a strong PT with $\Delta\varepsilon \sim \mathcal{O}(10^2) ~\mathrm{MeV}/\mathrm{fm}^3$.
We admit that the exact value is unavoidably model dependent --- it could rely on the based EoS, the number and distribution of the NS measurements, and the critical pressure at which the PT occurs.
Note that the quantitative analysis in Sec.~\ref{sec:results:md} has already taken $\sim 8\times10^5\,\mathrm{cpu}\cdot\mathrm{hours}$; the model dependence shall be analyzed in a neater way in the future.

Meanwhile, according to the sensitivity analysis, the tidal deformability turns out to be more sensitive to the EoS than the NS mass and radius, and we expect that possible future measurements of the $M$-$R$-$\Lambda$ relation shall be able to better constrain the EoS and detect the possible QCD phase transition. Explorations of the PT reconstruction using potential tidal deformability measurements of NSs will be performed in the future.
It is worth noting that the linear response analysis also applies to other observables of NSs, such as the moment of inertia and quadrupole moment. 
Such analyses could shed lights on the understanding of the approximate universal relations among these quantities.

Last but not least, this method can be easily recombined with other prior physics knowledge, such as chiral effective field theory~\cite{Drischler:2020yad, Drischler:2020hwi} and perturbative QCD~\cite{Komoltsev:2021jzg, Gorda:2022jvk, Gorda:2021znl} calculations, as well as constraints inferred from terrestrial nuclear experiments~\cite{CREX:2022kgg, PREX:2021umo, LeFevre:2015paj, Russotto:2016ucm, Danielewicz:2002pu, OmanaKuttan:2022aml}. 
It would be useful to reconstruct the dense matter EoS by combining all prior knowledge and all available astronomical measurements. 

\section*{Acknowledgments} 
The authors thank Zoey Zhiyuan Dong, Christian Drischler, Philippe Landry, Lap-Ming Lin, Shriya Soma, Andrew Steiner, and Boyang Sun for helpful discussions. 
We gratefully acknowledge the DEEP-IN working group at RIKEN-iTHEMS for support in the preparation of this paper. 
This work is supported by the National Key Research and Development Program of China under Contract No. 2024YFA1610700 (S.S.), Tsinghua University under grant No. 043-04200500123, No. 043-531205006, and No. 043-533305009 (S.S.), 
Guangdong Major Project of Basic and Applied Basic Research under grant No. 2020B030103000 (R.L.), 
Startup Funds from the T.D. Lee Institute and Shanghai Jiao Tong University (S.H.), 
NSF under grant No. PHY 21-16686 (Z.L.), 
the RIKEN TRIP initiative (RIKEN Quantum) and JST-BOOST grant (L.W.), the CUHK-Shenzhen university development fund under grant No. UDF01003041 and No. UDF03003041, and Shenzhen Peacock fund under No. 2023TC0179 (K.Z.).

\section*{Data Availability} 
The data that support the findings of this article are openly available~\cite{shi_2025_15086456}.

\begin{appendix}
\section{Numerical Validation of the Linear Response Analysis}
\label{app:validation}

\begin{figure*}[!htp]\centering
    \includegraphics[width=0.32\textwidth]{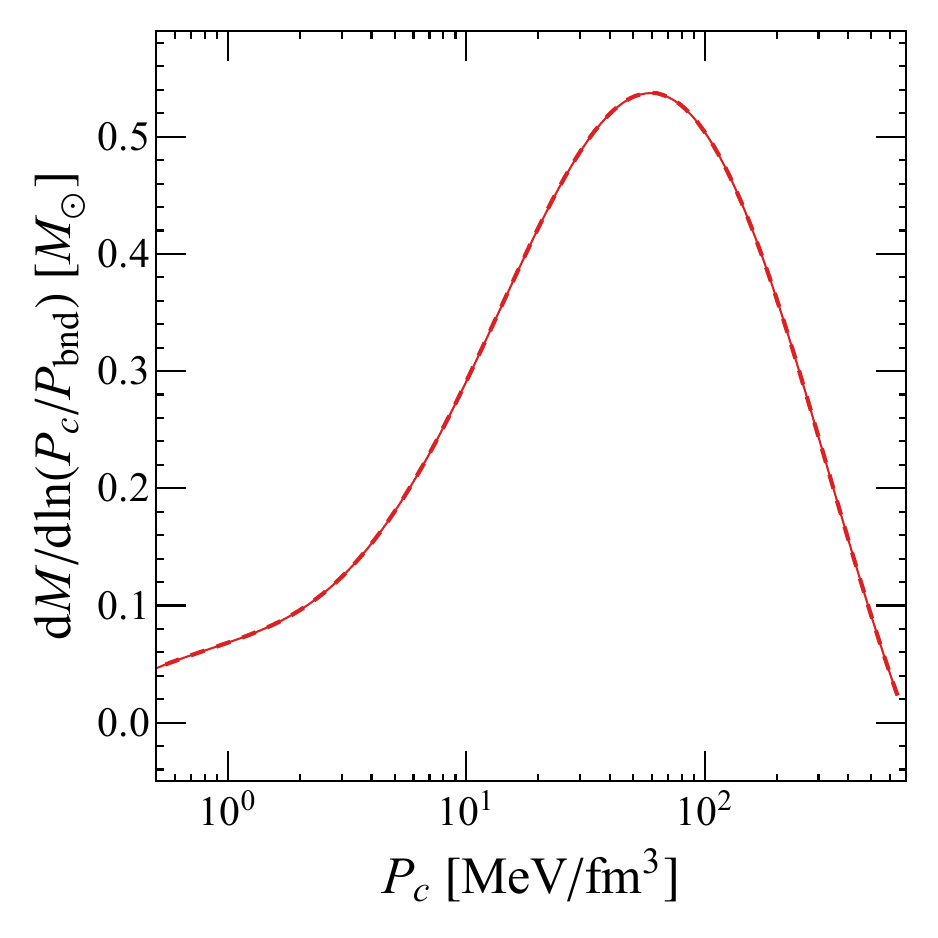}
    \includegraphics[width=0.32\textwidth]{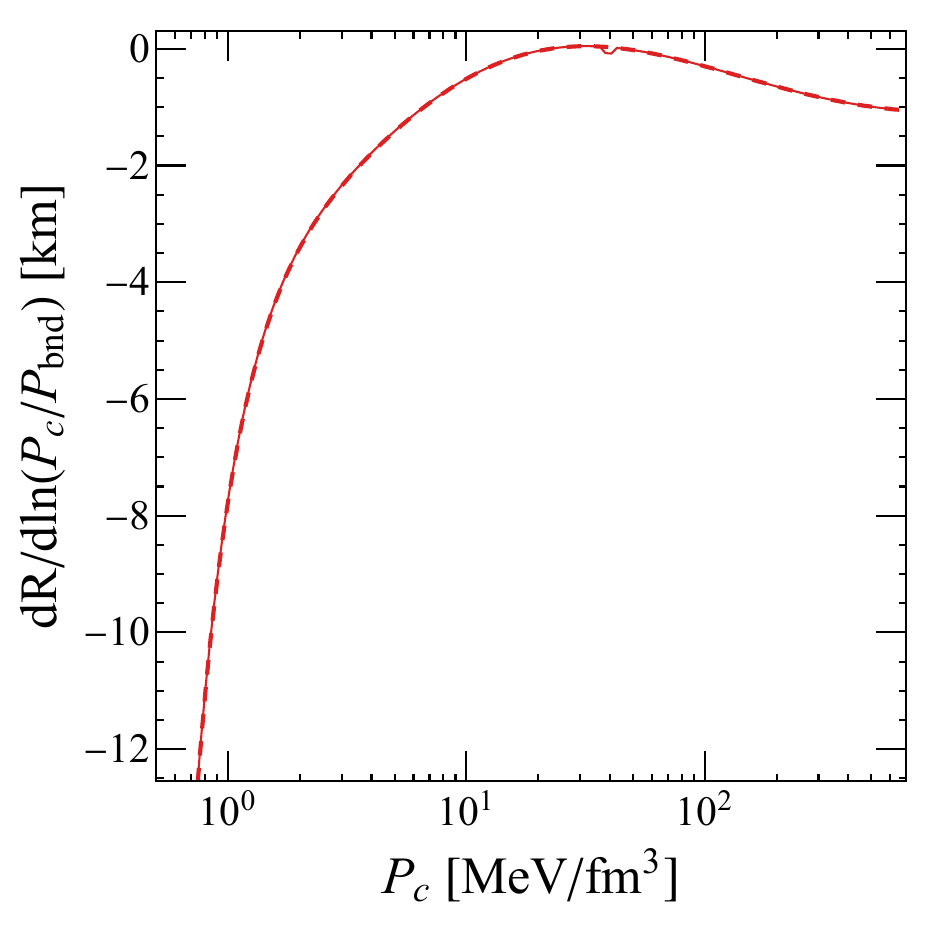}
    \includegraphics[width=0.32\textwidth]{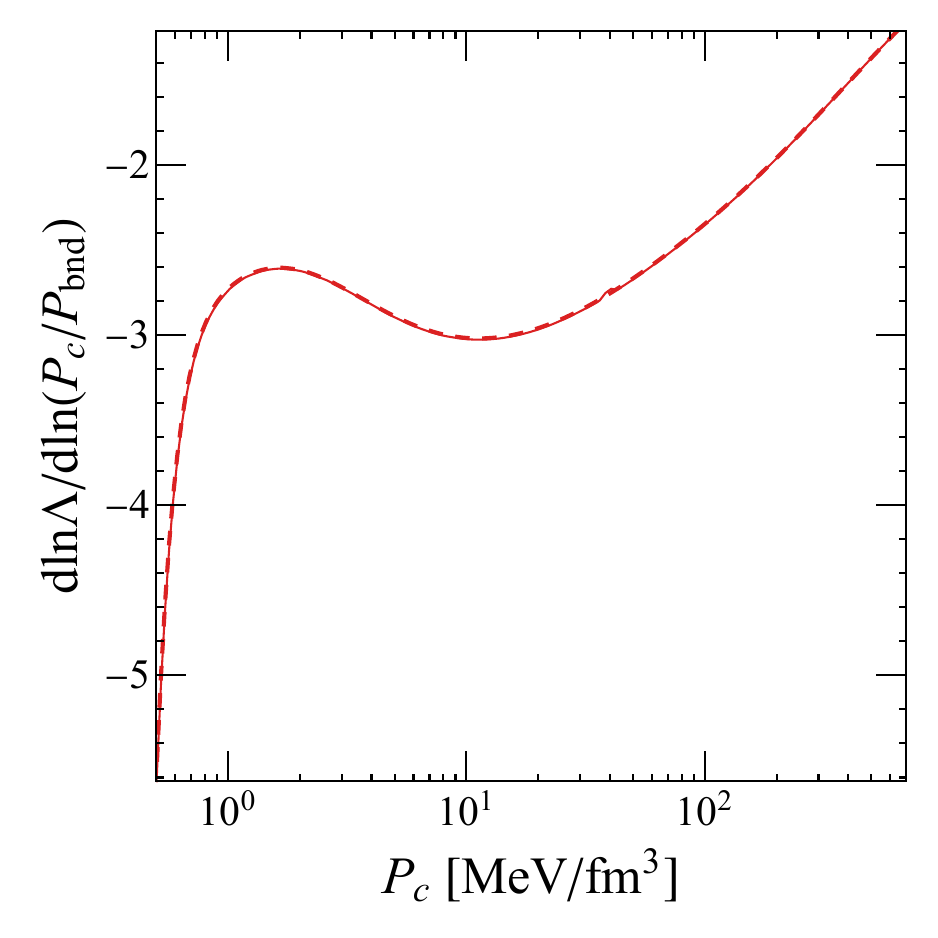}
    \includegraphics[width=0.32\textwidth]{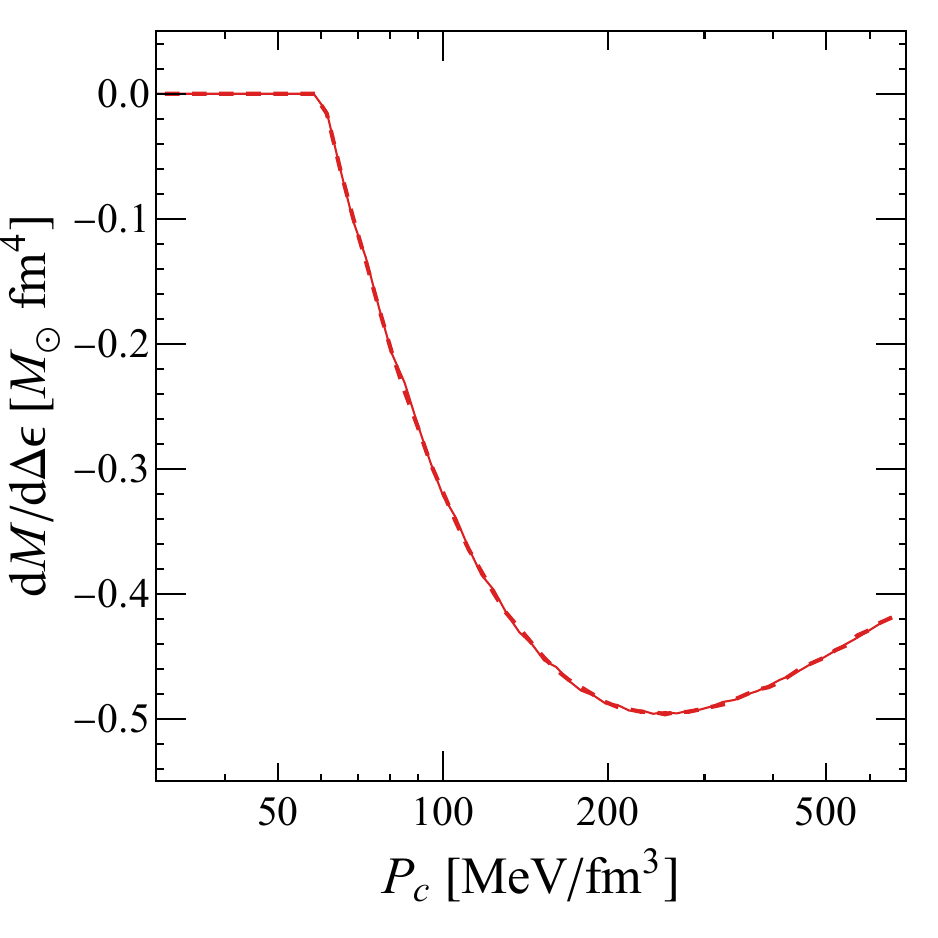}
    \includegraphics[width=0.32\textwidth]{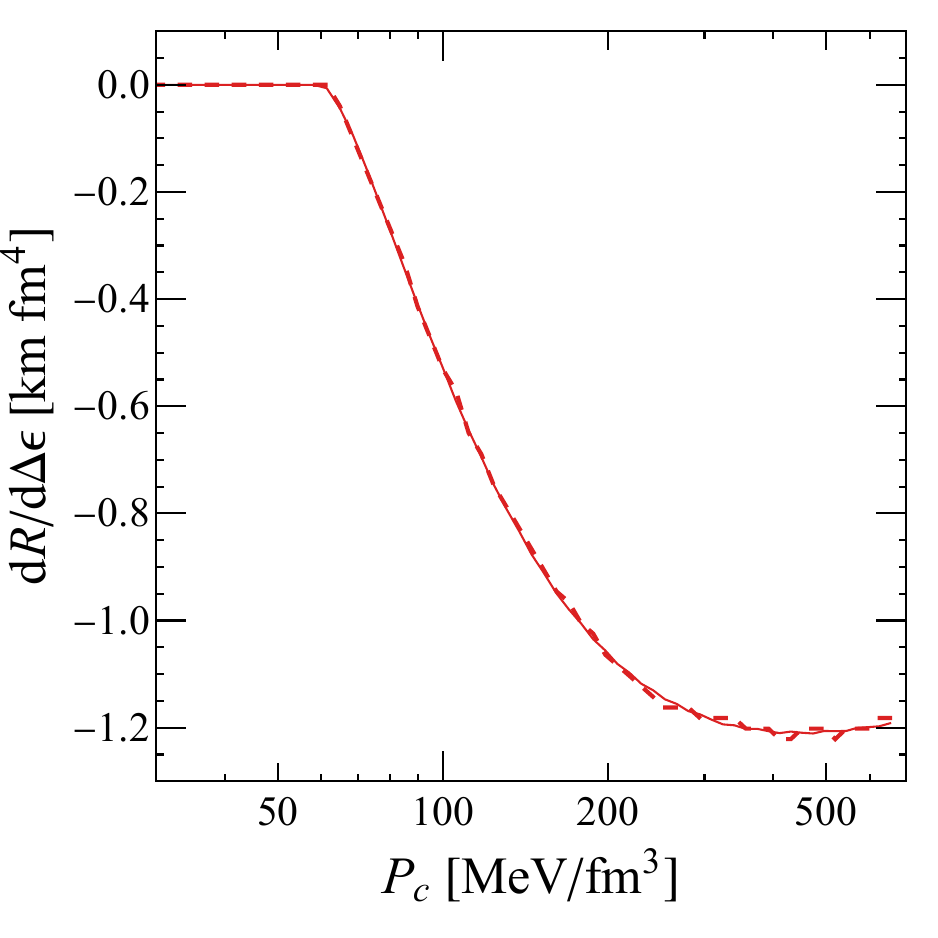}
    \includegraphics[width=0.32\textwidth]{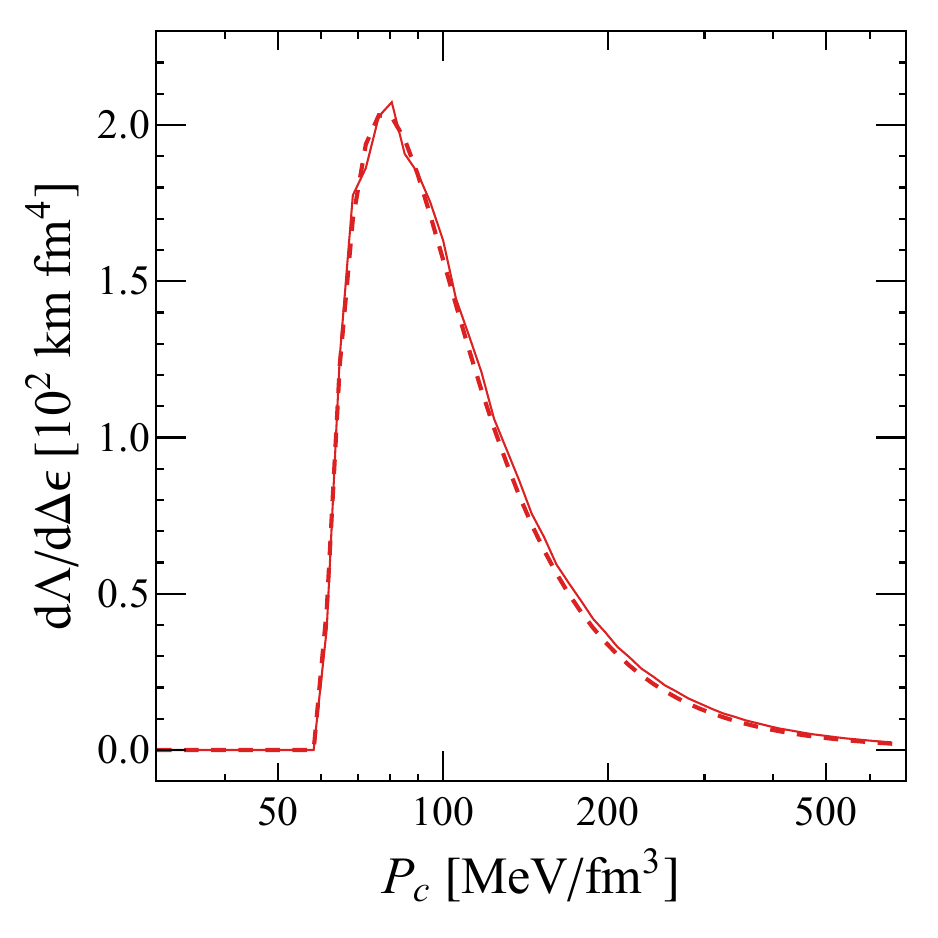}
    \caption{Change in the mass (left), radius (mid), and tidal deformability (right) against a perturbation in the central pressure (top) or a first-order phase transition (bottom).
    Solid curves represent results using formulae derived in \protect{Sec.~\ref{sec:algorithm}}, whereas dashed curves correspond to numerical derivatives taking finite differences.}
    \label{fig:devtest}
\end{figure*}

In this Appendix, we numerically validate the derivatives computed in Sec.~\ref{sec:algorithm}. We start from the \texttt{SFHo} EoS, compute the NS mass, radius, and tidal deformability for various central pressures (\{$P_c$\}). The obtained lists are denoted as $\{M\}$, $\{R\}$, $\{\Lambda\}$. To compute the numerical derivatives with respect to the central pressure, we repeat the procedure for a list of slightly different central pressures, $\{P_c\times(1+\delta\ln \frac{P_c}{P_\text{bnd}})\}$, and the corresponding observables are $\{\widetilde{M}\}$, $\{\widetilde{R}\}$, $\{\widetilde{\Lambda}\}$. 
Numerical derivatives are then computed as
$\{\frac{\widetilde{M}-M}{\delta\ln(P_c/P_\text{bnd})}\}$, $\{\frac{\widetilde{R}-R}{\delta\ln(P_c/P_\text{bnd})}\}$, and $\{\frac{\widetilde{\Lambda}-\Lambda}{\Lambda\delta\ln(P_c/P_\text{bnd})}\}$, 
which are compared to the results computed using Eqs.~(\ref{eq.dX_dPc_diff}, \ref{eq.dev_m_P}--\ref{eq.dev_Lambda}) 
and shown in the top panels of Fig.~\ref{fig:devtest}. $\delta\ln \frac{P_c}{P_\text{bnd}}$ is taken to be $\sim 0.04$ in numerical calculations.

We then examine the derivations of functional derivatives with respect to the EoS function. In order to do so, we obtain the lists $\{\overline{M}\}$, $\{\overline{R}\}$, and $\{\overline{\Lambda}\}$ according to the same central pressures, \{$P_c$\}, but with a phase transition included in the EoS,
\begin{align}
    \overline\varepsilon(P) = \varepsilon_{\texttt{SFHo}}(P) + \Delta \varepsilon \,\Theta(P-P_\mathrm{PT}). 
\end{align}
Numerical derivatives are then computed as
$\{\frac{\overline{M}-M}{\Delta\varepsilon}\}$, $\{\frac{\overline{R}-R}{\Delta\varepsilon}\}$, and $\{\frac{\overline{\Lambda}-\Lambda}{\Delta\varepsilon}\}$, which are compared to the results computed using Eqs.~(\ref{eq:eom:dv_de_diff}--\ref{eq:eom:dy_de_diff}, \ref{eq.dev_m_P}--\ref{eq.dev_Lambda}) and shown in the bottom panels of Fig.~\ref{fig:devtest}. 
In numerical calculations, we take $P_\mathrm{PT}=60~\text{MeV/fm}^3$ and $\Delta\varepsilon_\mathrm{PT}=1~\text{MeV/fm}^3$.

\section{Importance Sampling in Bayesian Analysis}
\label{app:importance_sampling}
\subsection{Definition of the Problem}

In Bayesian Analysis, with parameters denoted as $\boldsymbol{\vartheta}$ and their posterior distribution denoted as $\mathrm{P}(\boldsymbol{\vartheta})$, one needs to sample an ensemble of parameter sets ($\{\boldsymbol{\vartheta}_\ell\}_{\ell=1}^{L}$) that satisfies the posterior distribution. 
One can then approximate the posterior-weighted expectation of a given function of the parameters, denoted as $f(\boldsymbol{\vartheta})$, as the average of their values at the elements of such an ensemble,
\begin{align}
    \int f(\boldsymbol{\vartheta}) \mathrm{P}(\boldsymbol{\vartheta}) \,\mathrm{d}^N \boldsymbol{\vartheta} \approx \frac{1}{L}\sum_{\ell=1}^{L} f(\boldsymbol{\vartheta}_\ell)\,.
    \label{eq:target_expectation}
\end{align}
Here, $N$ is the dimension of the parameters, and $L$ is the number of the samples.

\subsection{Method I: Markov Chain Monte Carlo}

In traditional practice of Bayesian Analysis, one typically evoke the Markov Chain Monte Carlo (MCMC) to sample the parameter ensemble ($\{\boldsymbol{\vartheta}_\ell\}_{\ell=1}^{L}$). In MCMC, the parameter ensemble is sampled one after another as a chain. Given a parameter set in the chain, e.g., $\boldsymbol{\vartheta}_\ell$, the precedent set, $\boldsymbol{\vartheta}_{\ell+1}$, is sampled according to the following procedure:
\begin{itemize}
    \item[i)] Propose a new parameter set, $\tilde{\boldsymbol{\vartheta}}_{\ell+1}$, randomly.
    \item[ii)] Keep the proposed set with its probability being $\min\big(1, \mathrm{P}(\tilde{\boldsymbol{\vartheta}}_{\ell+1})/\mathrm{P}(\boldsymbol{\vartheta}_{\ell})\big)$. If kept, let $\boldsymbol{\vartheta}_{\ell+1}=\tilde{\boldsymbol{\vartheta}}_{\ell+1}$ and move on to sample $\boldsymbol{\vartheta}_{\ell+2}$; otherwise, repeat i).
\end{itemize}

In practice, there have been many methods to propose $\tilde{\boldsymbol{\vartheta}}_{\ell+1}$ in order to enhance the rate of acceptance, and one may need to drop a good portion of sets in the ensemble to remove correlation between the samples. 
Yet, no matter what accelerating methods are being used, sampling in a high dimensional parameter space would be computationally expensive and makes MCMC impractical.

\subsection{Method II: Importance Sampling}

When the parameter dimension is large, efficiency can be improved.
An alternative way would be to propose the samples according to a reference distribution that is easy to sample directly, $P_\mathrm{ref}(\boldsymbol{\vartheta})$, and then compute the target expectation~\eqref{eq:target_expectation} according to the reference distribution with weights being the ratio of two distributions,
\begin{align}
\begin{split}
    &\int f(\boldsymbol{\vartheta}) \mathrm{P}(\boldsymbol{\vartheta}) \,\mathrm{d}^N \boldsymbol{\vartheta} 
\\=\;&   
    \frac{\int \big(f(\boldsymbol{\vartheta}) \frac{\mathrm{P}(\boldsymbol{\vartheta})}{P_\mathrm{ref}(\boldsymbol{\vartheta})}\big) P_\mathrm{ref}(\boldsymbol{\vartheta}) \,\mathrm{d}^N \boldsymbol{\vartheta}}{\int \big(\frac{\mathrm{P}(\boldsymbol{\vartheta})}{P_\mathrm{ref}(\boldsymbol{\vartheta})}\big) P_\mathrm{ref}(\boldsymbol{\vartheta}) \,\mathrm{d}^N \boldsymbol{\vartheta}}
\\\approx\;&
    \frac{\sum_{\ell=1}^{L} f(\bar{\boldsymbol{\vartheta}}_\ell) \frac{\mathrm{P}(\bar{\boldsymbol{\vartheta}}_\ell)}{P_\mathrm{ref}(\bar{\boldsymbol{\vartheta}}_\ell)}}
    {\sum_{\ell=1}^{L} \frac{\mathrm{P}(\bar{\boldsymbol{\vartheta}}_\ell)}{P_\mathrm{ref}(\bar{\boldsymbol{\vartheta}}_\ell)}}\,,
\end{split}
\end{align}
where $\bar{\boldsymbol{\vartheta}}_\ell$'s are sampled according to the reference distribution, so that
\begin{align}
    \int f(\boldsymbol{\vartheta}) P_\mathrm{ref}(\boldsymbol{\vartheta}) \,\mathrm{d}^N \boldsymbol{\vartheta}
\approx
    \frac{1}{L}\sum_{\ell=1}^{L} f(\bar{\boldsymbol{\vartheta}}_\ell) \,.
\end{align}
The advantage is clear, that all $\bar{\boldsymbol{\vartheta}}_\ell$'s are sampled independently, and there is no correlation between them.

In such method, numerical efficiency would be highest if the weights $w_\ell \equiv \frac{\mathrm{P}(\boldsymbol{\vartheta}_\ell)}{P_\mathrm{ref}(\boldsymbol{\vartheta}_\ell)}$ are approximately unity. Otherwise, if $w_\ell$'s are different by orders of magnitude, the effective number of sample would be reduced, because the average would be dominated by samples with huge weights, and those with small weights become ``useless''. Out of the $L$ samples, we can estimate the effective number according to
\begin{align}
    L_\mathrm{eff} \equiv \frac{(\sum_{\ell=1}^{L} w_\ell)^2}{\sum_{\ell=1}^{L} w_\ell^2}.
\end{align}

In practice, if we manage to 1) find the parameter set ($\boldsymbol{\vartheta}^\mathrm{opt}$) that maximizes the posterior $\mathrm{P}(\boldsymbol{\vartheta})$ and 2) obtain the covariance matrix $C_{kk'} \equiv -\frac{1}{2}\frac{\partial^2 \ln \mathrm{P}(\boldsymbol{\vartheta})}{\partial \vartheta_k \partial \vartheta_{k'}}\big|_{\boldsymbol{\vartheta} = \boldsymbol{\vartheta}^\mathrm{opt}}$, then we may make our best guess of the reference distribution as the correlated Gaussian distribution around the optimal parameter set,
\begin{align}
    \mathrm{P}_\mathrm{ref}(\boldsymbol{\vartheta}) = \mathcal{N}_0 \, e^{- \frac{1}{2}C_{kk'}(\vartheta_k-\vartheta_{k}^\mathrm{opt})(\vartheta_{k'}-\vartheta_{k'}^\mathrm{opt})}.
    \label{eq:reference_distribution}
\end{align}

\subsection{Marginal Posterior Distribution using Importance Sampling}
\label{app:importance_sampling:mpd}

In the main text, we are particularly interest in the marginal posterior distribution of the pressure and latent heat of the first-order phase transition, which is defined as
\begin{align}
    \mathrm{P}(\xi_\mathrm{PT}, \Delta \varepsilon) \equiv 
    \int \mathrm{P}(\bth) \mathrm{d}^{N-2}\hat\bth\,,
\end{align}
where $\hat\bth$ denotes parameters other than $\xi_\mathrm{PT}$ and $\Delta \varepsilon$. It should be noted that the overall posterior is normalized, $1=\int \mathrm{P}(\bth) \mathrm{d}^{N-2}\hat\bth \,\mathrm{d}\xi_\mathrm{PT}  \,\mathrm{d}\Delta \varepsilon$, so that the marginal one is also normalized, $\int \mathrm{P}(\xi_\mathrm{PT}, \Delta \varepsilon) \,\mathrm{d}\xi_\mathrm{PT}  \,\mathrm{d}\Delta \varepsilon$.

For a better estimation of $\mathrm{P}(\xi_\mathrm{PT}, \Delta \varepsilon)$, we select $\sim 200$ points in the pressure and latent heat parameter space --- labeled as $\xi_{\mathrm{PT},j}$ and $\Delta \varepsilon_j$, respectively --- and optimized other parameters (denoted as $\hat\bth_j^\mathrm{opt}$) correspondingly. 
We then sample $\hat\bth$'s according to the reference distribution
\begin{align}
P_{\mathrm{ref},j}(\hat\bth) = 
    \mathcal{N}_j e^{-\frac{1}{2} (\hat\bth-\hat\bth^\mathrm{opt}_j)^T\cdot \boldsymbol{C}_j \cdot(\hat\bth-\hat\bth^\mathrm{opt}_j)},
\end{align}
and compute
\begin{align}
\mathrm{P}(\xi_{\mathrm{PT},j}, \Delta \varepsilon_j) 
\approx \frac{1}{L}\sum_{\ell=1}^{L} \frac{\mathrm{P}(\xi_{\mathrm{PT},j}, \Delta \varepsilon_j, \hat\bth_\ell)}{P_{\mathrm{ref},j}(\hat\bth_\ell)}\,.
\end{align}
The uncertainty of $\mathrm{P}(\xi_{\mathrm{PT},j}, \Delta \varepsilon_j)$ can be estimated by the variance of weights, $w_{\ell}^{(j)}\equiv\frac{\mathrm{P}(\xi_{\mathrm{PT},j}, \Delta \varepsilon_j, \hat\bth_\ell)}{P_{\mathrm{ref},j}(\hat\bth_\ell)}$, i.e., $\delta \mathrm{P}(\xi_{\mathrm{PT},j}, \Delta \varepsilon_j) = \big(\sum_{\ell=1}^{L} (w_{\ell}^{(j)})^2 / L - (\sum_{\ell=1}^{L}w_{\ell}^{(j)}/L)^2\big)^{\frac{1}{2}}/L^{\frac{1}{2}}$, which also equals $\mathrm{P}(\xi_{\mathrm{PT},j}, \Delta \varepsilon_j)/(L_\mathrm{eff}^{(j)})^{\frac{1}{2}}$.

With $\mathrm{P}(\xi_{\mathrm{PT},j}, \Delta \varepsilon_j)$ and $\delta \mathrm{P}(\xi_{\mathrm{PT},j}, \Delta \varepsilon_j)$ obtained for all points on the grid, we further invoke Gaussian Process to estimate $\mathrm{P}(\xi_{\mathrm{PT}}, \Delta \varepsilon)$ for the region of interest, and correspondingly construct the credible region.

\end{appendix}

\bibliographystyle{apsrev4-1}
\bibliography{references}
\end{document}